\documentclass[]{aastex6}

\newcommand{\kzz}{$K_{zz}$}

\begin{document}


\title{Sedimentation Efficiency of Condensation Clouds in Substellar Atmospheres}



\author{Peter Gao\altaffilmark{1,2}}
\affil{University of California, Berkeley \\
Berkeley, CA 94720, USA}

\author{Mark S. Marley}
\affil{NASA Ames Research Center \\
Moffett Field, CA 94035, USA}

\and

\author{Andrew S. Ackerman}
\affil{NASA Goddard Institute for Space Studies \\
2880 Broadway \\
New York, NY 10025, USA}




\altaffiltext{1}{51 Pegasi b Fellow}
\altaffiltext{2}{gaopeter@berkeley.edu}

\begin{abstract}

Condensation clouds in substellar atmospheres have been widely inferred from spectra and photometric variability. Up until now, their horizontally averaged vertical distribution and mean particle size have been largely characterized using models, one of which is the eddy diffusion--sedimentation model from \citet{ackerman2001} that relies on a sedimentation efficiency parameter, $f_{\rm sed}$, to determine the vertical extent of clouds in the atmosphere. However, the physical processes controlling the vertical structure of clouds in substellar atmospheres are not well understood. In this work, we derive trends in $f_{\rm sed}$ across a large range of eddy diffusivities ($K_{zz}$), gravities, material properties, and cloud formation pathways by fitting cloud distributions calculated by a more detailed cloud microphysics model. We find that $f_{\rm sed}$ is dependent on $K_{zz}$, but not gravity, when $K_{zz}$ is held constant. $f_{\rm sed}$ is most sensitive to the nucleation rate of cloud particles, as determined by material properties like surface energy and molecular weight. High surface energy materials form fewer, larger cloud particles, leading to large $f_{\rm sed}$ ($>$1), and vice versa for materials with low surface energy. For cloud formation via heterogeneous nucleation, $f_{\rm sed}$ is sensitive to the condensation nuclei flux and radius, connecting cloud formation in substellar atmospheres to the objects' formation environments and other atmospheric aerosols. These insights could lead to improved cloud models that help us better understand substellar atmospheres. For example, we demonstrate that $f_{\rm sed}$ could increase with increasing cloud base depth in an atmosphere, shedding light on the nature of the brown dwarf L/T transition.

\end{abstract}




\section{Introduction} \label{sec:intro}

Clouds have been readily inferred from observations of exoplanet and brown dwarf atmospheres. In exoplanet transmission spectroscopy, optically thick clouds block photons from reaching below the cloud tops, resulting in flat transmission spectra or diminutive molecular features \citep[e.g.][]{kreidberg2014,sing2016}. In reflected light, clouds lead to increased albedos as compared to clear atmospheres and shifted bright spots at visible wavelengths \citep[e.g.][]{demory2013,parmentier2016}. In thermal emission, cloud opacity decreases the depth of molecular absorption bands in brown dwarf spectra and increases the observed day--night contrast in tidally locked exoplanets, while patchy clouds lead to temporal variability in brown dwarf photometry \citep[e.g.][]{tsuji1996,allard2001,tsuji2002,stephens2009,helling2014review,stevenson2017,biller2017}. These effects reflect the importance of accounting for clouds in analyses of exoplanet and brown dwarf observations, and the treatment of clouds in retrievals can result in different conclusions on temperature structure and composition \citep{benneke2015,line2016}.  

An oft-used cloud model in interpreting exoplanet and brown dwarf data is that of \citet{ackerman2001}, hereupon known as AM01. It calculates the 1--D cloud mass and particle size distributions by balancing sedimentation with lofting due to eddy diffusion, under the simplifying assumption that these clouds are horizontally homogeneous. The vertical extent of the cloud is controlled by the $f_{\rm sed}$ parameter, which sets the efficiency with which particles can settle out of the cloud. Small $f_{\rm sed}$ values ($f_{\rm sed}<$1) correspond to low efficiency, and thus vertically extended clouds with small particles, while large $f_{\rm sed}$ values ($f_{\rm sed}>$1) point to more vertically compressed clouds with large particles. AM01 has been applied to interpret a variety of exoplanet and brown dwarf observations \citep[e.g.][]{fortney2006,mainzer2007,stephens2009,cushing2010,burgasser2011,buenzli2012,heinze2013,gelino2014,morley2015,esplin2016,rajan2017}. It has also been expanded to treat patchy clouds \citep{marley2010} and been used to create large model grids for retrievals \citep{skemer2016,henderson2017}. Throughout these works the value of $f_{\rm sed}$ has ranged from as small as 0.01 for super Earths \citep{morley2015} to nearly 10 for some brown dwarfs \citep{saumon2008}. However, because the value of $f_{\rm sed}$ is determined solely from the quality of the fit to observations, the physical processes that lead to such $f_{\rm sed}$ values are not known. In other words, it is not apparent why $f_{\rm sed}$ should be small in some cases while large in others, or whether the $f_{\rm sed}$ values determined from the observations can physically manifest from the chosen atmospheric and condensate characteristics. 

The size and spatial distribution of cloud particles in planetary atmospheres is controlled by cloud microphysical processes occurring on the scale of the cloud particles that lead to particle formation (nucleation), growth (condensation, coagulation), and loss (evaporation, breakup). Simulations of these processes require more complex and time-consuming models than AM01, but they can physically motivate the choice of $f_{\rm sed}$, thereby shedding light on the role of clouds in previous and future exoplanet and brown dwarf atmospheric studies. Indeed, simple and complex cloud models are complementary; while AM01 can be used as part of iterative radiative--convective models, retrievals, and creating model atmosphere grids that require hundreds of simulations, more complex cloud microphysics models can offer detailed insights that AM01 cannot. However, comparisons between simple and complex cloud models are seldom done. 

One example of a complex cloud microphysics model is DRIFT \citep{helling2008c,helling2013}. It has been used primarily to simulate brown dwarf and hot Jupiter atmospheres \citep[e.g.][]{witte2011,lee2016,lee2017} by treating cloud formation as a kinetic process, beginning with TiO$_{2}$ clusters forming by homogeneous nucleation \citep{lee2015b} high up in the atmosphere, and upon sedimentation act as condensation nuclei for a range of species simultaneously, including forsterite, enstatite, iron, corundum, quartz, etc., resulting in mixed ``dirty grains''. The condensate mass mixing ratio, mean cloud particle size, cloud composition, and other quantities computed by DRIFT are compared to those computed by AM01 and several other models in \citet{helling2008a}, which showed large differences in cloud properties calculated by the different models owing to the different assumptions made about the cloud formation process (kinetic versus phase--equilibrium). However, \citet{helling2008a} did not attempt to reproduce AM01 results using DRIFT or characterize $f_{\rm sed}$ in its comparison between the two models. In this work, we compare the results of a more general cloud microphysics model to that of AM01 to find how $f_{\rm sed}$ varies with different atmospheric and condensate properties. We describe the AM01 and microphysics models in ${\S}$\ref{sec:models}, along with the test cases we use to compare them. We describe our results in ${\S}$\ref{sec:results}, where we show how the cloud particle size and spatial distribution and $f_{\rm sed}$ vary with the eddy diffusivity \kzz, gravity, condensate surface energy, and formation pathway. We discuss the implications of our results in the context of observed trends in exoplanet and brown dwarf cloudiness in ${\S}$\ref{sec:discussion}, and state our conclusions in ${\S}$\ref{sec:conclusions}.

\section{Models}\label{sec:models}

\subsection{Ackerman \& Marley (2001)}

AM01 calculates the molar mixing ratio of condensed material, $q_c$, by solving 

\begin{equation}
\label{eq:am01}
-K_{zz}\frac{\partial q_t}{\partial z} - f_{\rm sed} \omega_{*} q_c = 0
\end{equation}

\noindent where $q_t$ is the total condensate molar mixing ratio (including both the condensed and vapor phases), $z$ is altitude, and $\omega_{*}$ is the convective velocity scale. Eq. \ref{eq:am01} states that the upward mixing of condensate vapor and particles is balanced by the downward sedimentation of particles. Heuristically, under the assumption that $q_c/q_t$ and \{the mixing length, $L$, are constant with altitude and given a $q_t$ value below the cloud, $q^{\rm below}_t$, the $q_t$ profile above the cloud base is given by

\begin{equation}
\label{eq:qtcloud}
q_t(z) = q^{\rm below}_t \exp \left (-f_{\rm sed} \frac{q_c}{q_t}\frac{z}{L} \right )
\end{equation}

\noindent with all supersaturated condensate vapor condensing and $z$ = 0 at the cloud base. \{In the full model, $L$ is defined as the ratio of the local lapse rate to the dry adiabatic lapse rate times the scale height $H$, with a lower limit of 0.1$H$. The eddy diffusivity $K_{zz}$ is then related to $L$ by 

\begin{equation}
\label{eq:kzz}
K_{zz} = \frac{H}{3} \left ( \frac{L}{H} \right )^{4/3} \left ( \frac{RF_h}{\mu_a \rho_a C_p} \right )^{1/3}
\end{equation}

\noindent where $R$ is the universal gas constant, $\mu_a$ is the atmospheric molecular weight, $\rho_a$ is the atmospheric mass density, and $C_p$ is the specific heat capacity of the atmosphere at constant pressure. $F_h$ is the local heat flux carried by convection, which approaches the interior heat flux $\sigma T_{\rm eff}^4$ in the deep, convective interior of a planet, where $\sigma$ is the Stefan--Boltzmann constant and $T_{\rm eff}$ is the effective temperature. In radiative regions, $K_{zz}$ is set to a specified minimum value. For simplicity, we set $K_{zz}$ to be a constant throughout the atmosphere in our study (see ${\S}$\ref{sec:modelsetup}). The convective velocity scale $\omega_{*}$ is then given by $K_{zz}/L$.  
 
The particle size distribution is assumed to be lognormal, with the effective (area-weighted) particle radius given by 

\begin{equation}
\label{eq:reff}
r_{\rm eff} = r_g  \exp \left ( \frac{5}{2} \ln^2 \sigma_g \right )
\end{equation}

\noindent where $\sigma_g$ is the geometric standard deviation of the lognormal distribution, set to 2 to crudely span the condensation and coagulation modes of cloud particles, and $r_g$ is the geometric mean particle radius, defined through integration of the lognormal distribution by

\begin{equation}
\label{eq:rg}
r_g = r_w f_{\rm sed}^{1/\alpha} \exp \left (- \frac{\alpha + 6}{2} \ln^2 \sigma_g \right )
\end{equation}

\noindent The system is closed analytically by locally fitting the dependence of the particle radius $r$ on the sedimentation velocity of particles $v_f$ using a power law around the radius $r_w$ such that $v_f(r_w)$ = $\omega_{*}$

\begin{equation}
\label{eq:rwalpha}
v_f = w_{*} \left ( \frac{r}{r_w} \right )^{\alpha}
\end{equation}
 
\noindent with $v_f$ given for Stokes flow by  

 \begin{equation}
\label{eq:vf}
v_f = \frac{2}{9}\frac{\Delta \rho g r^2 \beta}{\eta}
\end{equation}

\noindent where $\Delta \rho$ is the difference in density between the background atmosphere and the density of the particle, $g$ is the gravitational acceleration, $\beta$ = 1 + 1.26Kn is the Cunningham slip correction factor, with Kn the Knudsen number of the particle, and $\eta$ is the atmospheric dynamic viscosity. At Reynolds numbers between 1 and 1000, $v_f$ is augmented by a standard parameterization with respect to the drag coefficient $C_d$ and Reynolds number, while at Reynolds numbers greater than 1000, $v_f$ is given by

\begin{equation}
\label{eq:vf1000}
v_f = \beta \sqrt{\frac{8 \Delta \rho g r}{3 C_d \rho_a}}
\end{equation}

\subsection{CARMA}

We use the Community Aerosol and Radiation Model for Atmospheres (CARMA) as the ``standard'' to which we compare AM01. CARMA is a 1--dimensional aerosol microphysics model that solves the discretized continuity equation for aerosol particles subject to vertical transport due to sedimentation and eddy diffusion and production and loss due to particle nucleation (homogeneous and heterogenous), condensation, evaporation, and coagulation. \{CARMA was initially used to model Earth's stratospheric sulfate aerosols \citep{turco1979,toon1979}, and has since been generalized to a variety of applications both on Earth and in other planetary atmospheres, including modeling polar stratospheric clouds to inform ozone depletion \citep{toon1988}, the characteristics of the particles stemming from the eruption of Mount Pinatubo \citep{zhao1995}, various tropospheric cloud features on Earth \citep{ackerman1993,jensen1994carma,ackerman1995}, the sulfuric acid clouds of Venus \citep{james1997,mcgouldrick2007,gao2014}, water ice clouds on Mars \citep{colaprete1999}, and photochemical hazes on titan \citep{toon1992}, Pluto \citep{gao2017pluto}, and ancient Earth \citep{wolf2010}. Extension of the model to 3 dimensions \citep{toon1989} has allowed for the study of Martian dust storms \citep{murphy1993} and meteoric dust in Earth's mesosphere \citep{bardeen2008}, among others. In the rest of this section we describe qualitatively CARMA's treatment of the various cloud processes and we refer the reader to Appendix \ref{sec:carma} for a full description of the physics that is included in CARMA.

CARMA discretizes the vertical extent of the atmosphere into layers and resolves the particle size distribution using mass bins rather than assuming a particular size distribution shape. The sedimentation velocity is computed in the same way as in AM01, but the particle velocities are calculated for each mass bin individually. Similarly, the diffusion velocity associated with eddy diffusion (given a user-defined $K_{zz}$) is calculated for particles in each mass bin and for condensate vapors according to \citet{toon1988}. Both the sedimentation and diffusion velocities are then combined to solve for the vertical particle distribution at each time step. 

\{The formation of clouds in CARMA begins with homogeneous or heterogeneous nucleation. Homogeneous nucleation is the formation of stable clusters of condensate molecules directly from the vapor that can then grow to larger cloud particles. The rate of homogeneous nucleation is controlled by (1) the flux of molecules to the cluster, which is dependent on the abundance of condensate vapor, and (2) the material properties of the condensate, such as its surface energy and molecular weight. High surface energy and molecular weight materials tend to nucleate slower than low surface energy and molecular weight materials given the same supersaturation and local temperature. Unlike homogeneous nucleation, heterogeneous nucleation requires foreign surfaces on which stable clusters can form; such foreign surfaces are provided by other aerosol particles in the atmosphere, typically described as condensation nuclei. Thus, the size and abundance of these particles strongly impact the rate of heterogeneous nucleation. In addition, the nucleation rate is dependent on the interaction between the condensate and the surface, characterized by the contact angle between the condensate cluster and the surface, the energy needed by a condensate molecule to desorb from said surface, and the oscillation frequency of the condensate molecule on said surface, which is related to the desorption energy \citep{pruppacher1978}. The values of the contact angle and the desorption energy are unknown for the substances considered here and must be estimated. Typical contact angles for water range from $\sim$0$^{\circ}$ for highly hydrophilic surfaces and $>$100$^{\circ}$ for highly hydrophobic surfaces \citep[e.g.][]{ethington1990,yuan2013}; small contact angles lead to high heterogeneous nucleation rates and large contact angles lead to low heterogeneous nucleation rates. As our study is focused on trends in cloud distribution instead of simulating any specific known system, we use a small contact angle (0.1$^{\circ}$) for all our heterogeneous nucleation cases to increase the efficiency of cloud formation and the optical depth of the resulting cloud. Thermal desorption energies for simple molecules (e.g. H$_{2}$O, CO$_{2}$, CH$_{4}$, etc.) over silicate grains in the interstellar medium are typically on the order of 0.1 eV \citep{seki1983,suhasaria2015,suhasaria2017}, though energies $>$1 eV are also possible for metallic surfaces \citep[e.g. desorption of potassium from a nickel surface;][]{blaszczyszyn1995}. We adopt a desorption energy of 0.18 eV and corresponding oscillation frequency of 10$^{13}$ Hz for that of water over silicate \citep{seki1983} for all of our simulations. While the desorption energy could potentially be higher, it would not strongly impact our results as the magnitude of the nucleation rate depends more on the exponential term \{in the rate equation (Eq. \ref{eq:hetnuceq}).

\{Following nucleation, cloud particles can grow via condensation or shrink by evaporation. The condensation and evaporation rates in CARMA are partially determined, like nucleation, by the flux of condensate molecules towards or away from the surface of cloud particles. In addition, growth may be limited by the rate with which particles can conduct away the latent heat released upon condensation, while the opposite process can be applied to limit evaporation. Particles can also grow via coagulation, where cloud particles physically stick to each other upon collision and grow larger at the expense of decreasing number densities. For simplicity we do not consider coagulation in this work, as it is not clear whether exoplanet and brown dwarf particles would coagulate to form larger spheres or aggregates \citep{lavvas2017}. If the former, then the resulting larger particles would increase the corresponding $f_{\rm sed}$ value of the cloud. In contrast, formation of low density aggregates may result in clouds that are highly extended vertically, decreasing $f_{\rm sed}$. We will investigate these effects in a future publication.

\subsection{Model Setup}\label{sec:modelsetup}

We use a set of exoplanet and brown dwarf atmospheres to conduct our model comparisons. These atmospheres originate from a larger model grid detailed in Marley et al. (in prep), and are composed of temperature-pressure-composition profiles in radiative-convective-thermochemical equilibrium, assuming no external insolation and H-He-dominated atmospheres with solar metallicity. 

\begin{figure}[hbt!]
\centering
\includegraphics[width=0.6 \textwidth]{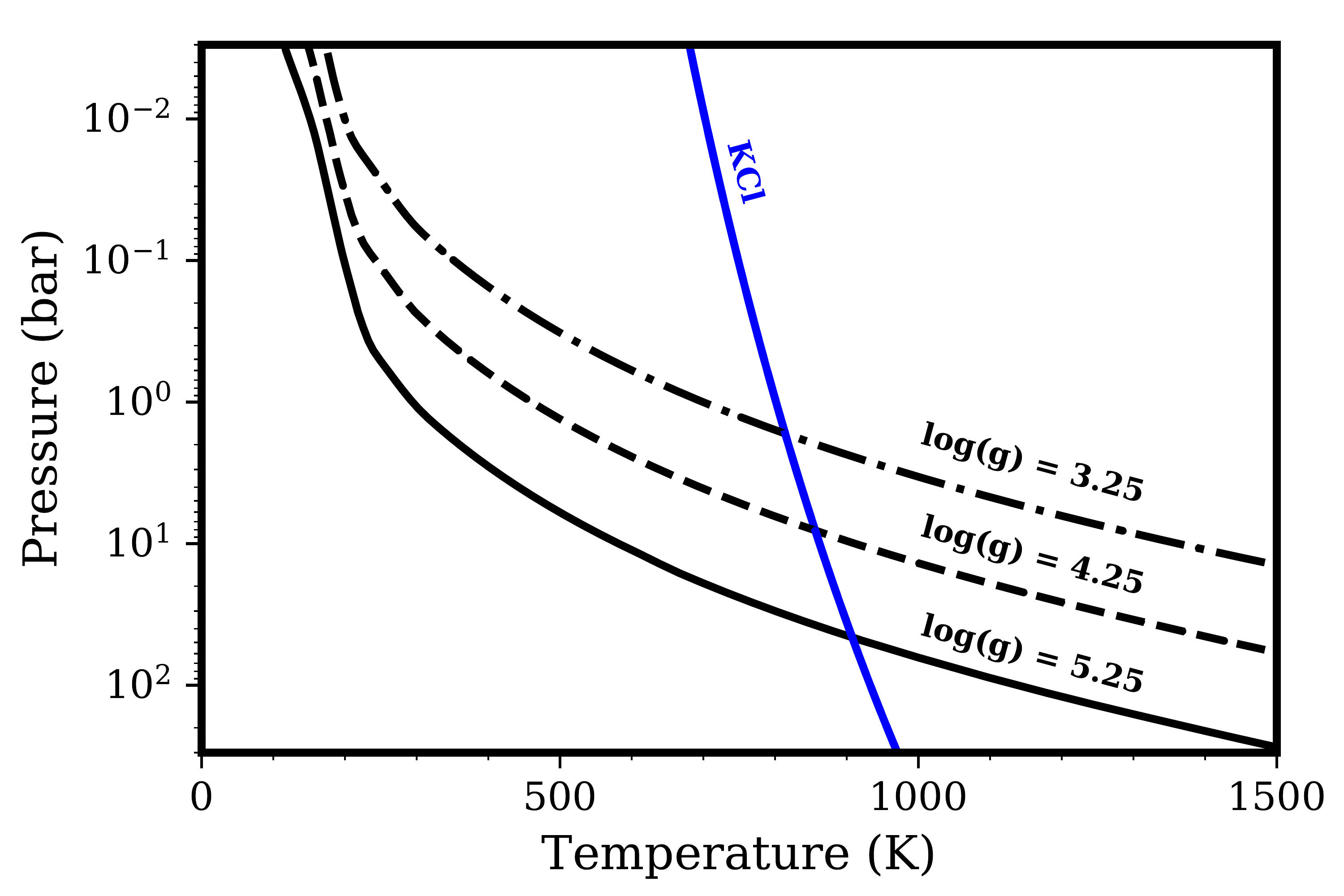}
\caption{Pressure--temperature profiles of the model atmospheres considered in this work, corresponding to an effective temperature of 400 K and log $g$ = 3.25 (dash-dot), 4.25 (dashed), and 5.25 (solid). KCl's condensation curve is shown in blue.}
\label{fig:tp}
\end{figure}

We consider objects with an effective temperature of 400 K and log $g$ = 3.25, 4.25, and 5.25 ($g$ in cgs units), corresponding to masses of 0.72M$_{\rm J}$, 8.47M$_{\rm J}$, and 44.54M$_{\rm J}$, respectively. \{Fig. \ref{fig:tp} shows the pressure--temperature profiles of our model atmospheres, which cover the pressure range from 64 bars to 0.18 mbar. For the log $g$ = 5.25 case we also test 3 separate constant-with-altitude $K_{zz}$ profiles with values of 10$^{6}$, 10$^{7}$, and 10$^{8}$ cm$^{2}$ s$^{-1}$ to assess the impact of changing $K_{zz}$ on the value of $f_{\rm sed}$. As $K_{zz}$ is calculated from Eq. \ref{eq:kzz} in AM01 and we are replacing it with a constant profile in our tests, we use Eq. \ref{eq:kzz} to calculate $L$  given our constant $K_{zz}$ profiles to ensure self-consistency. The effective temperature of 400 K was chosen to allow for the formation of potassium chloride (KCl) clouds, which is the only type of cloud we consider in this study. KCl is an optimal choice because it undergoes simple phase transition between vapor and solid or liquid upon cloud formation, rather than relying on chemical reactions \citep{morley2012} that could complicate the cloud nucleation process. Note that objects with an effective temperature of 400 K and log $g$ = 5.25 do not yet exist, as objects of such high mass takes longer than the age of the Universe to cool to such low temperatures. This is acceptable in the present study since we are only interested in determining the general trend in the cloud distribution, rather than predicting the actual cloudiness of real objects. 

Given the reliance of the nucleation rate on the surface energy of the condensate and, for heterogeneous nucleation, the size and abundance of condensation nuclei, we also conduct two additional sets of tests -- one to investigate how $f_{\rm sed}$ of the KCl cloud changes with different KCl surface energies, and one to investigate how the size and downward flux of condensation nuclei affect the $f_{\rm sed}$ of the resulting KCl cloud. For the surface energy tests we alter the surface energy of KCl as a function of temperature $T$ \citep{westwood1963,janz1969}

\begin{equation}
\label{eq:kclsurftens}
\sigma^{\rm KCl}_s(\rm ergs\, cm^{-2}) = 160.4 - 0.07T(\rm K)
\end{equation}

\noindent by decreasing it by factors of 2 and 4 and increasing it by factors of 2, 3, and 4. For the heterogeneous nucleation tests we assume condensation nuclei radii of 0.1, 1, and 10 nm, and downward fluxes of 10, 100, and 1000 cm$^{-2}$ s$^{-1}$, assuming the condensation nuclei to be composed of silicates with density $\sim$2 g cm$^{-3}$, similar to the meteoritic dust generated by ablating meteorites in Earth's atmosphere. By comparison, Earth receives 100--300 metric tons of interplanetary dust particles per day \citep{plane2012}, which corresponds to a flux of 27000--81000 cm$^{-2}$ s$^{-1}$ of 1 nm particles, though only a fraction of this mass would be capable of acting as condensation nuclei. The full list of test cases are given in Table \ref{table:all}. 

Comparison of the two models are accomplished by first running CARMA to equilibrium. CARMA is initialized with a model atmosphere devoid of KCl vapor or cloud particles. KCl vapor is then allowed to diffuse upwards from depth, given a fixed lower boundary mixing ratio of 0.22 ppmv, \{corresponding to the total abundance of K in a solar metallicity atmosphere \citep{lodders2010}. Thus, we assume that all K is locked in KCl, which is valid given the pressure--temperature space occupied by the KCl clouds in our model upon their formation \citep[see Fig. \ref{fig:tp} in this work and Fig. 2 in][]{lodders1999}. Upon reaching saturation, with the saturation vapor pressure estimated by \citet{morley2012} to be

\begin{equation}
\label{eq:kclsvp}
\log{p_s^{\rm KCL} (\rm bars)} = 7.611 - 11382/T(\rm K)
\end{equation}

\noindent KCl nucleates homogeneously and form a cloud deck. \{Fig. \ref{fig:tp} compares our model pressure--temperature profiles to KCl's condensation curve; regions of the model atmospheres cooler than the condensation curve are amenable to KCl cloud formation. After nucleation, the KCl cloud particles are free to grow by condensation, shrink by evaporation, and be transported by sedimentation and diffusion. 

We set a zero--flux boundary condition at the top of the model domain in the homogeneous nucleation cases, while a finite flux of condensation nuclei is set for the heterogeneous nucleation cases. A total of 65 particle mass bins are used, with the smallest bin corresponding to particles with radius 1 \AA, and the mass doubling for successive bins. The time step sizes for the different cases were tuned to avoid numerical errors in the nucleation and condensation processes, which occur when the time step size is too large, and to avoid extremely long run times when the time step size is too small. Model run times are dominated by the mixing time scale, which is related to $K_{zz}$ and the scale height. Thus, cases with high $K_{zz}$ and/or high gravity were run using short time steps, and vice versa for cases with lower $K_{zz}$ and/or low gravity. The time step sizes are given in Table \ref{table:all}. Changing the time step size does not lead to any significant changes to the model results. The models were deemed converged once the column mass of particles and vapor became constant with time.

\subsection{Model Comparison}\label{sec:modelcomp}

We calculate $\epsilon q_c$ for the KCl clouds computed by CARMA as a function of pressure level by summing the masses of all particles within a pressure level and dividing it by the local atmospheric mass density, where $\epsilon$ is the ratio of the condensate molecular weight to the mean atmospheric molecular weight. $r_{\rm eff}$ at each pressure level is calculated by finding the weighted average particle radii, with the total cross sectional area of particles at that level as a function of particle radii acting as weights,

\begin{equation}
r_{\rm eff} = \frac{\int N_r r^3dr}{\int N_r r^2 dr}
\end{equation}

\noindent where $N_r$ is the number density of particles as function of particle radius. We combine $q_c$ and $r_{\rm eff}$ to calculate the optical depth $\Delta \tau$ per atmospheric layer $\Delta z$ for geometric scatterers by using (AM01)

\begin{equation}
\label{eq:dtau}
\Delta \tau = \frac{3}{2} \frac{\epsilon \rho_a q_c}{\rho_p r_{\rm eff}} \Delta z 
\end{equation}

\noindent where $\rho_p$ is the particle mass density. Summing $\Delta \tau$ from the top of the atmosphere downwards gives the cumulative optical depth. 

To find the optimal $f_{\rm sed}$ value that ``fit'' the cloud distributions computed by CARMA, we minimize the difference in the pressure level $P_{0.1}$ where the two cloud distributions from the two cloud models reach cumulative optical depths of 0.1. This is an optimal comparison criterion because the more diffuse, upper portions of the cloud do not strongly impact transmission, reflection, and emission observations, and thus it is irrelevant to assert that the two models are equal there. Though we do not assert that the two cloud distributions are similar at pressures greater than $P_{0.1}$, nor do we constrain $r_{\rm eff}$ in any way, we find that our comparison criterion ensures that these other quantities are very similar between the two models as well, unless the total cumulative optical depth of the cloud distribution computed by CARMA is less than 0.1. The actual ``fitting'' of $f_{\rm sed}$ is done by running the AM01 model 10000 times with $f_{\rm sed}$ increasing by steps of 0.001 each time from $f_{\rm sed}$ = 0.001 to 10, and comparing $P_{0.1}$ for the AM01 model with that of CARMA. The $f_{\rm sed}$ value corresponding to the minimum in difference of $P_{0.1}$ between the two models is deemed the $f_{\rm sed}$ that best ``fits'' the CARMA cloud distribution. 

\section{Results}\label{sec:results}

\subsection{Variations with $K_{zz}$ and Gravity}

\begin{deluxetable}{cccccccc}
\tablecolumns{8}
\tablecaption{Run parameters and best-fit $f_{\rm sed}$ values.\label{table:all}}
\tablehead{
\colhead{Case} & \colhead{Time Step (s)} & \colhead{$K_{zz}$ (cm$^{2}$ s$^{-1}$)} & \colhead{log $g$} & \colhead{$\sigma_s$\tablenotemark{a}} & \colhead{$r_n$ (nm)} & \colhead{CN\tablenotemark{b} Flux (cm$^{-2}$ s$^{-1}$)}  & \colhead{Best-Fit $f_{\rm sed}$}}
\startdata
1	&	1	&	$10^6$	&	5.25		&	$\sigma_s^{KCl}$	&	\nodata	&	\nodata	&	0.125	\\
2	&	0.1	&	$10^7$	&	5.25		&	$\sigma_s^{KCl}$	&	\nodata	&	\nodata	&	0.093	\\
3	&	0.01	&	$10^8$	&	5.25		&	$\sigma_s^{KCl}$	&	\nodata	&	\nodata	&	0.025	\\
4	&	1	&	$10^8$	&	4.25		&	$\sigma_s^{KCl}$	&	\nodata	&	\nodata	&	0.050	\\
5	&	100	&	$10^8$	&	3.25		&	$\sigma_s^{KCl}$	&	\nodata	&	\nodata	&	0.036	\\
6	&	0.1	&	$10^7$	&	5.25		&	0.25 $\times$ $\sigma_s^{KCl}$	&	\nodata	&	\nodata	&	0.036	\\
7	&	0.1	&	$10^7$	&	5.25		&	0.5 $\times$ $\sigma_s^{KCl}$	&	\nodata	&	\nodata	&	0.059	\\
8	&	0.1	&	$10^7$	&	5.25		&	2 $\times$ $\sigma_s^{KCl}$	&	\nodata	&	\nodata	&	0.150	\\
9	&	0.1	&	$10^7$	&	5.25		&	3 $\times$ $\sigma_s^{KCl}$	&	\nodata	&	\nodata	&	0.353	\\
10	&	0.1	&	$10^7$	&	5.25		&	4 $\times$ $\sigma_s^{KCl}$	&	\nodata	&	\nodata	&	$>$10	\\
11	&	0.01	&	$10^8$	&	5.25		&	$\sigma_s^{KCl}$	&	1	&	10	&	$>$10	\\
12	&	0.01	&	$10^8$	&	5.25		&	$\sigma_s^{KCl}$	&	1	&	100	&	7.745	\\
13	&	0.01	&	$10^8$	&	5.25		&	$\sigma_s^{KCl}$	&	1	&	1000		&	0.938	\\
14	&	0.01	&	$10^8$	&	5.25		&	$\sigma_s^{KCl}$	&	10	&	1000		&	0.937	\\
15	&	0.01	&	$10^8$	&	5.25		&	$\sigma_s^{KCl}$	&	0.1	&	1000		&	1.311	\\
\enddata
\tablenotetext{a}{See Eq. \ref{eq:kclsurftens} for the definition of $\sigma_s^{KCl}$.}
\tablenotetext{b}{Condensation Nuclei}
\end{deluxetable}

\begin{figure}[hbt!]
\centering
\includegraphics[width=1.0 \textwidth]{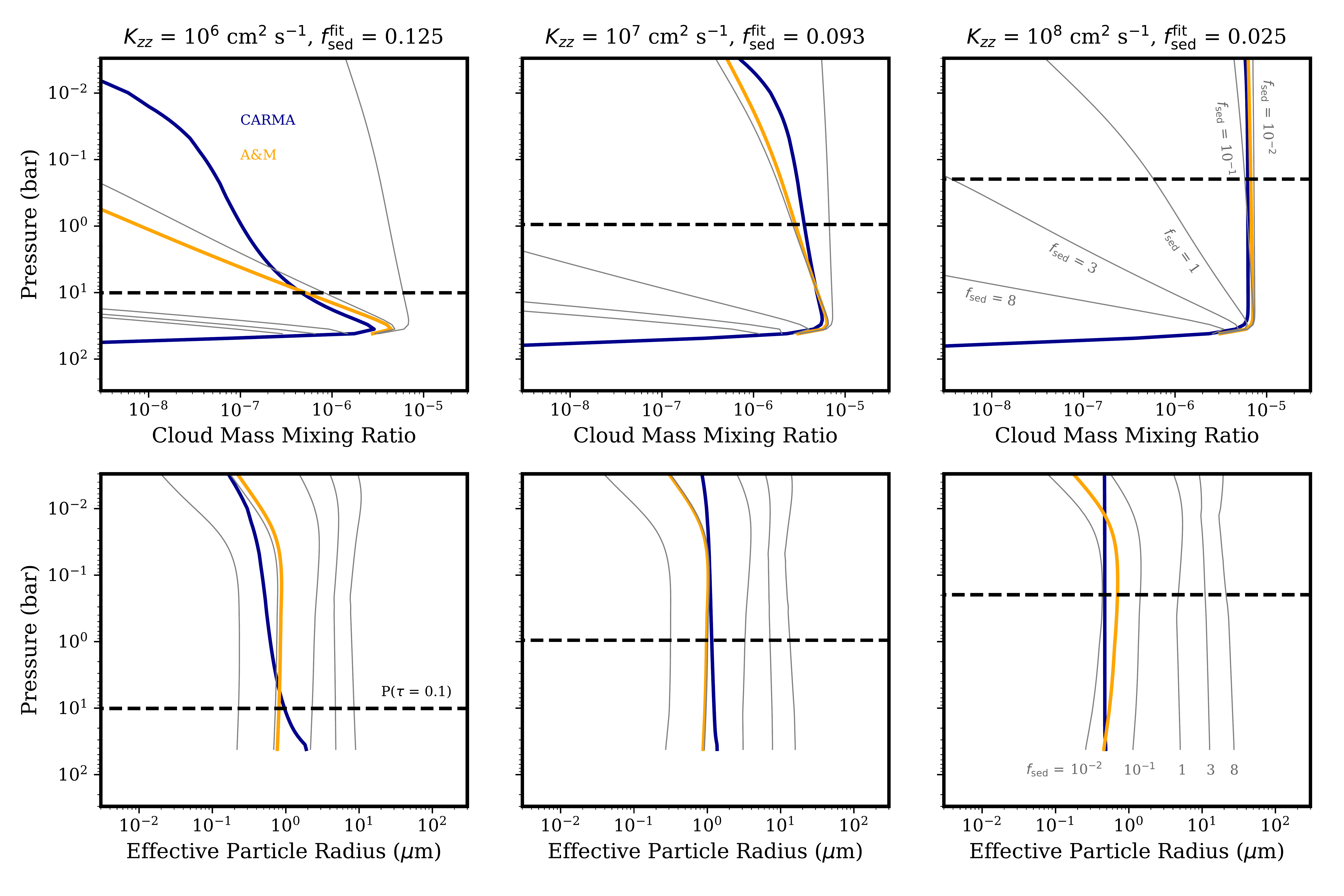}
\caption{Cloud mass mixing ratio $q_c$ (top) and cloud particle effective radius $r_{\rm eff}$ (bottom) profiles computed by CARMA (blue) compared to that of the best fit $f_{\rm sed}$ AM01 profiles (orange) as functions of atmospheric pressure level for the $K_{zz}$ = $10^6$ (left), $10^7$ (middle), and $10^8$ cm$^{2}$ s$^{-1}$ (right) cases. log $g$ is fixed to 5.25. $q_c$ and $r_{\rm eff}$ profiles corresponding to other $f_{\rm sed}$ values are presented for comparison. The pressure levels where the cumulative optical depth reaches 0.1 from above are shown as horizontal dashed lines. Only homogeneous nucleation is considered. }
\label{fig:kzzvar}
\end{figure}

We find that the best fit $f_{\rm sed}$ decreases with increasing $K_{zz}$ (Fig. \ref{fig:kzzvar}; Table \ref{table:all}). While increasing the mixing strength within the atmosphere should produce more extended clouds, our results show that microphysical processes serve to further enhance vertical extension. The increased mixing leads to higher fluxes of condensate vapor into the cloud forming region, resulting in increased rates of cloud particle nucleation and the formation of numerous small particles. This decreases the cloud's sedimentation efficiency, thereby creating a more vertically extended cloud. Though the cloud particles do grow by condensation from the upward flux of vapor, most of the growth is isolated to near the cloud base, where $r_{\rm eff}$ reaches a maximum for the CARMA clouds. This is not seen in the AM01 clouds, where $r_{\rm eff}$ is roughly constant with depth. The increased vertical extent of the cloud with increased $K_{zz}$ leads to an upward movement of the $\tau$ = 0.1 surface in the atmosphere, such that observations probe decreasing column densities and lower temperatures as $K_{zz}$ increases (assuming no thermal inversions). 

\begin{figure}[hbt!]
\centering
\includegraphics[width=1.0 \textwidth]{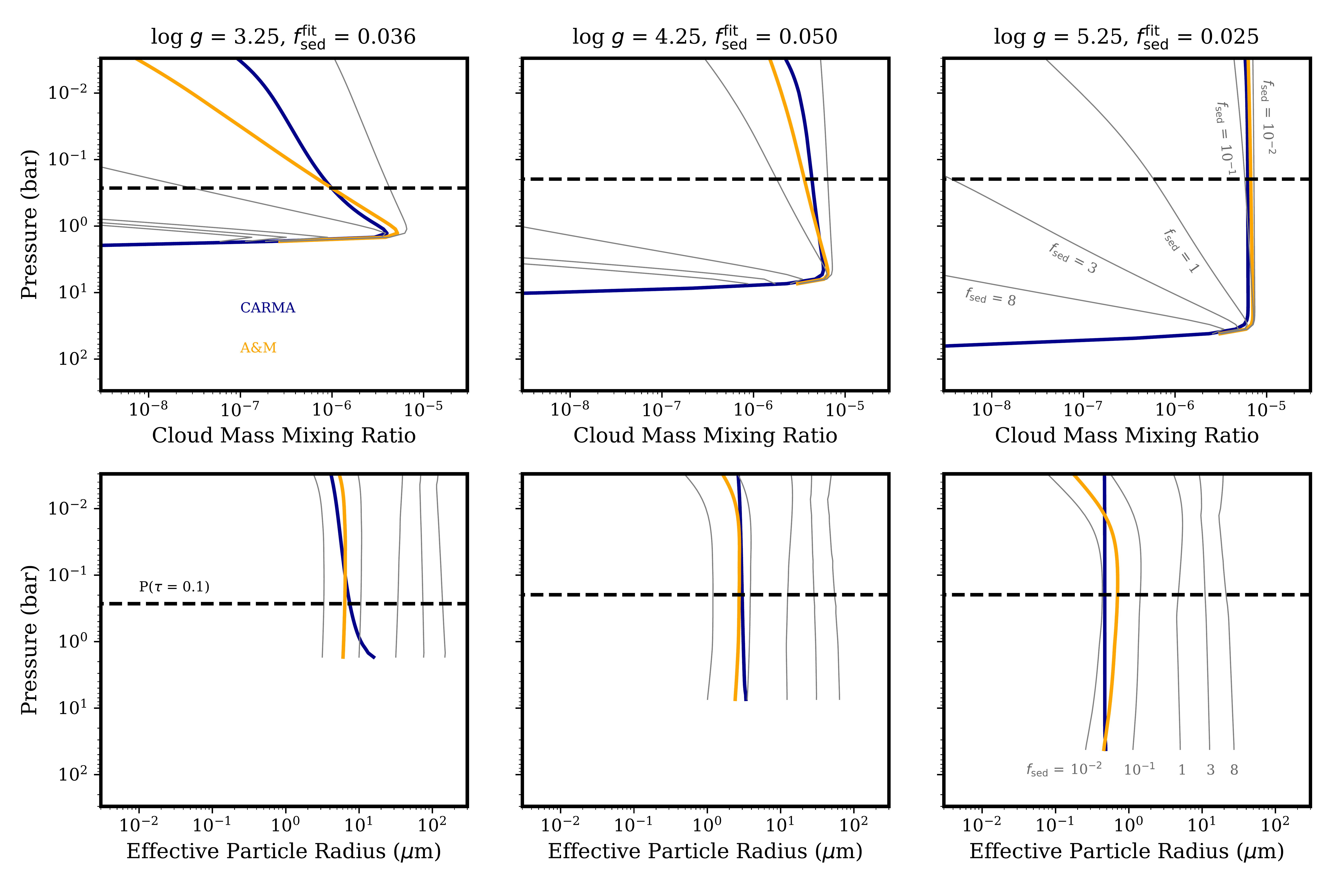}
\caption{ Same as Fig. \ref{fig:kzzvar}, but for log $g$ = 3.25 (left), 4.25 (middle), and 5.25 (right) cases. $K_{zz}$ is fixed to $10^8$ cm$^{2}$ s$^{-1}$. }
\label{fig:gravvar}
\end{figure}

Unlike with $K_{zz}$, the relationship between $f_{\rm sed}$ and gravity is non-monotonic (Fig. \ref{fig:gravvar}; Table \ref{table:all}). This can be understood as the action of multiple processes largely canceling each other out. For example, the sedimentation and mixing time scales of cloud particles can be estimated by 

\begin{equation}
\label{eq:sedtime}
t_{sed} = \frac{H}{v_f} \propto \frac{1}{g^2} 
\end{equation}

\begin{equation}
\label{eq:mixtime}
t_{mix} = \frac{H^2}{K_{zz}} \propto \frac{1}{g^2} 
\end{equation}

\noindent and therefore their combined effects cancel out, assuming constant $K_{zz}$. However, condensate vapor is not directly affected by sedimentation, and so the decrease in $t_{mix}$ at the highest log $g$ case could be due to increased vapor fluxes leading to higher rates of nucleation, as with the case with increased $K_{zz}$. In addition, due to lower log $g$ atmospheres having higher temperatures despite the same effective temperature (note for example the decreasing altitude of the cloud base with increasing log $g$), part of the (small) difference in $f_{\rm sed}$ could be due to temperature effects on the nucleation and growth rates of particles. For instance, lower temperatures may lead to lower rates of nucleation (though this could be compensated by higher saturation ratios), and therefore a slight increase in $f_{\rm sed}$ for the higher log $g$ case (4.25 vs. 3.25) may be expected. Curiously, the pressure level at which $\tau$ = 0.1 does not change considerably between the three log $g$ cases. Given that greater scale heights lead to higher column densities, observations should probe higher column densities and temperatures with decreasing gravity. 

\subsection{Variations with Surface Energy}\label{sec:varsurf}

\begin{figure}[hbt!]
\centering
\includegraphics[width=1.0 \textwidth]{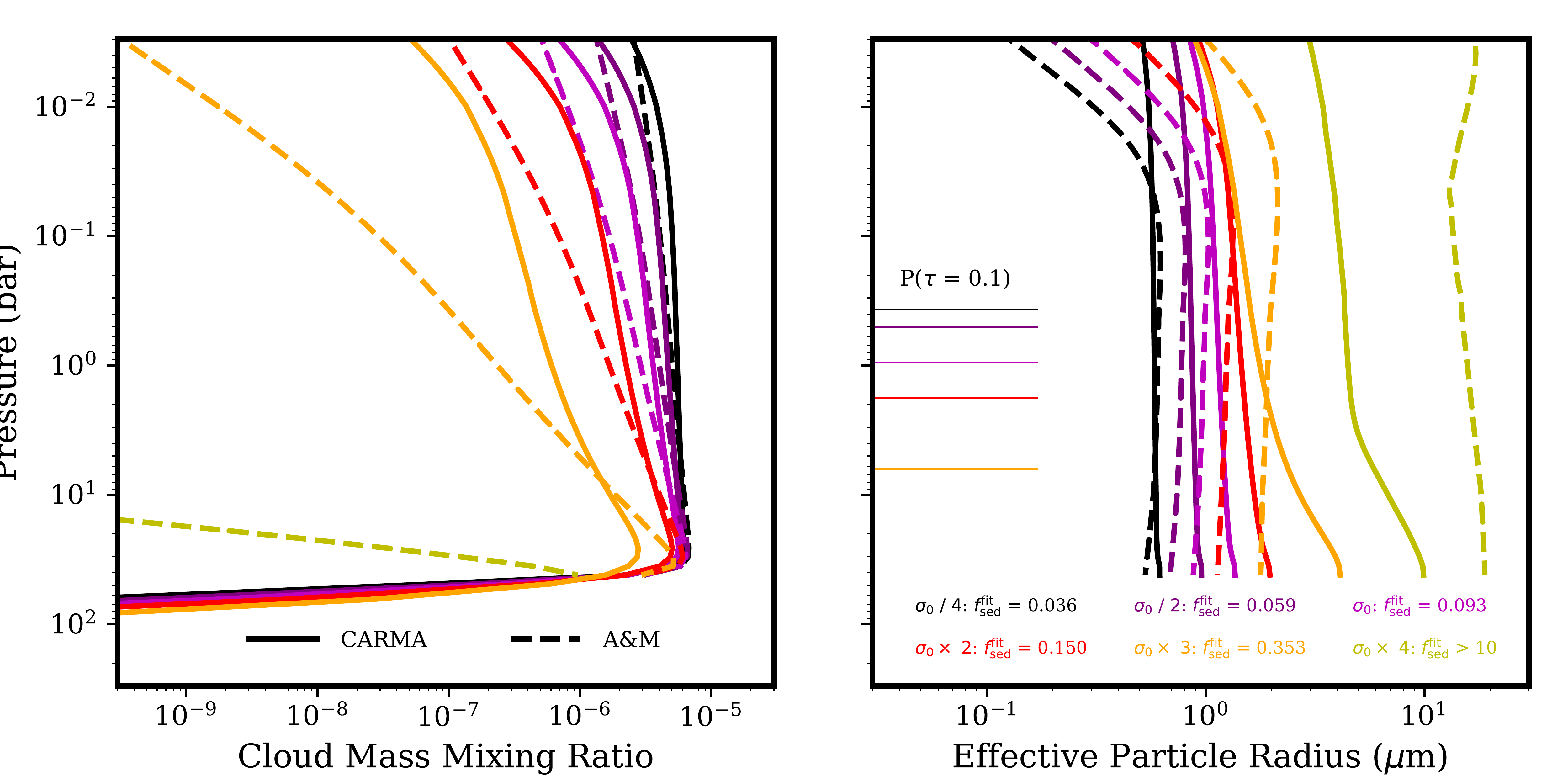}
\caption{Cloud mass mixing ratio $q_c$ (left) and effective cloud particle radius $r_{\rm eff}$ (right) profiles computed by CARMA (solid) compared to that of the best fit $f_{\rm sed}$ AM01 profiles (dashed) as functions of atmospheric pressure level for the test cases where KCl's surface energy is altered (see cases 6 to 10 in Table \ref{table:all}; case 3 is also plotted for comparison). $K_{zz}$ is fixed to $10^7$ cm$^{2}$ s$^{-1}$ and log $g$ is fixed to 5.25. The pressure levels where the cumulative optical depth reaches 0.1 from above are shown as horizontal solid lines with colors corresponding to the $q_c$ and $r_{\rm eff}$ profiles. The highest surface energy cloud never reaches a cumulative optical depth of 0.1 and in fact has a $q_c$ profile smaller than the lower limit of the plot on the left. Only homogeneous nucleation is considered. }
\label{fig:surfenvar}
\end{figure}

\begin{figure}[hbt!]
\centering
\includegraphics[width=0.6 \textwidth]{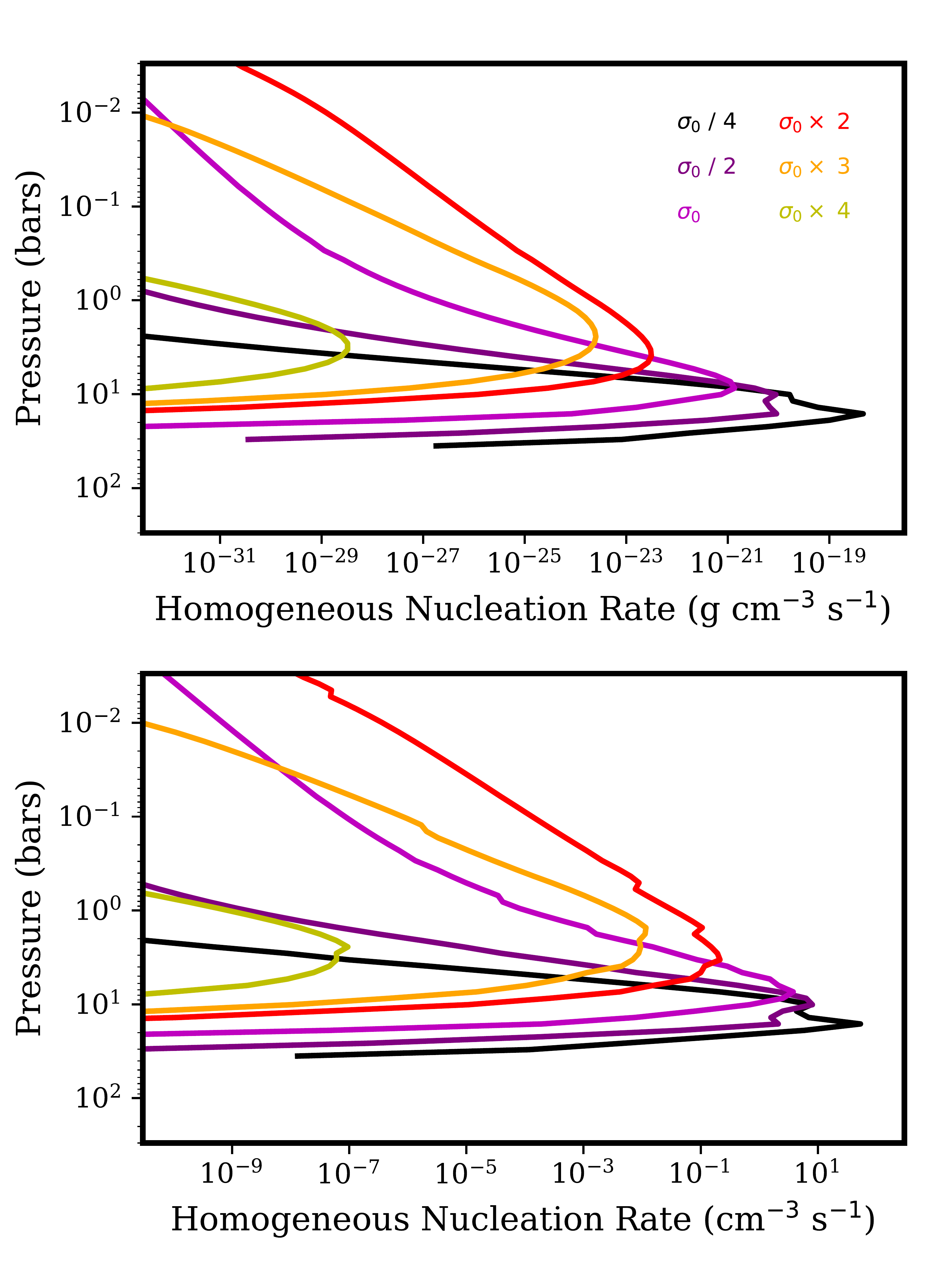}
\caption{Homogeneous nucleation rates in units of mass density (top) and particle number density (bottom) for the test cases where KCl's surface energy is altered. $K_{zz}$ is fixed to $10^7$ cm$^{2}$ s$^{-1}$ and log $g$ is fixed to 5.25. The ``steps'' in the curves in the bottom plot is due to how nucleating particles are allocated between the discrete particle mass bins and does not signficantly impact our results.}
\label{fig:homnucprofile}
\end{figure}

\begin{figure}[hbt!]
\centering
\includegraphics[width=0.6 \textwidth]{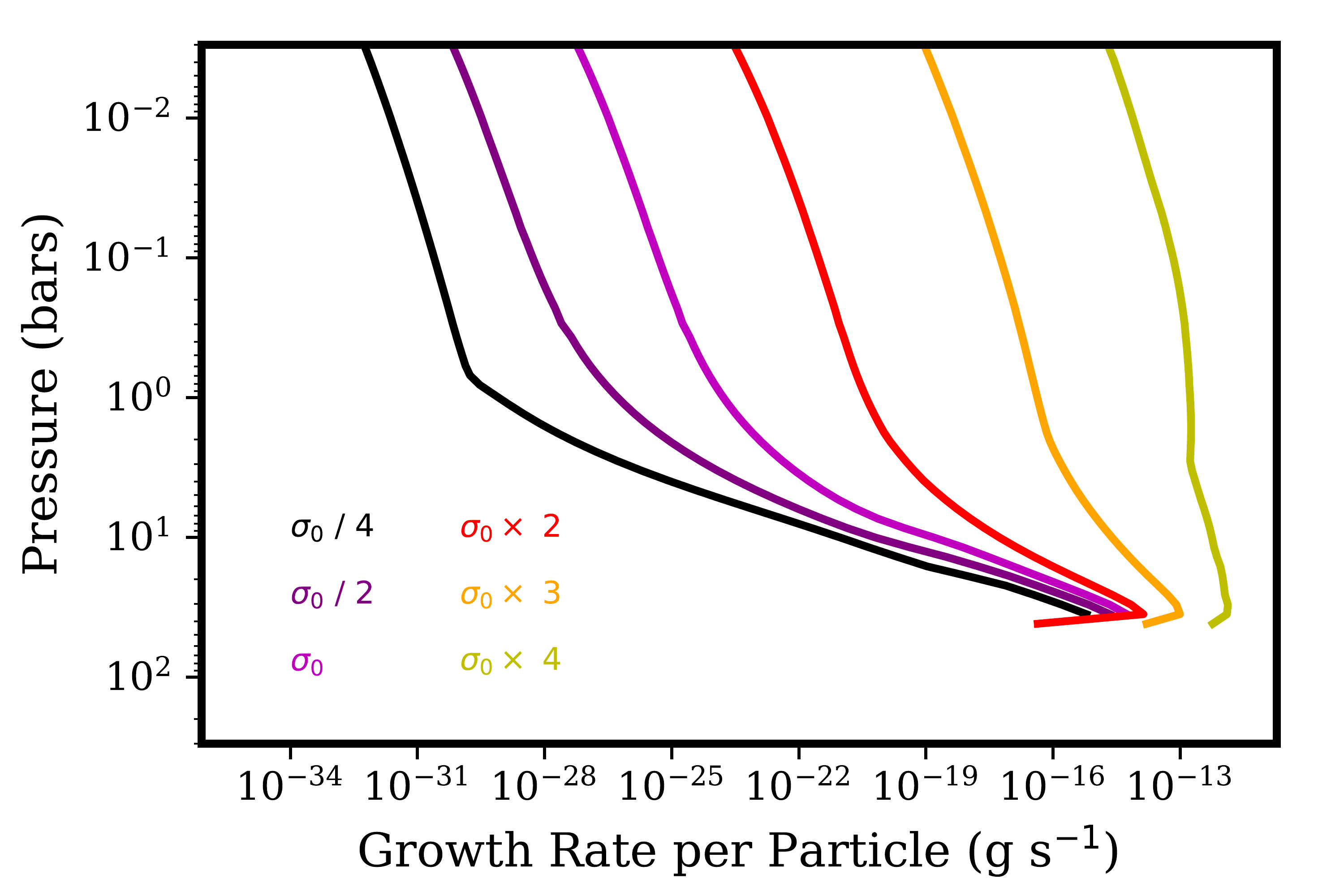}
\caption{Growth rate per cloud particle for the test cases where KCl's surface energy is altered. $K_{zz}$ is fixed to $10^7$ cm$^{2}$ s$^{-1}$ and log $g$ is fixed to 5.25.}
\label{fig:growparticle}
\end{figure}

The $f_{\rm sed}$ values for the cases where $K_{zz}$ and g are varied ($\sim$0.1) are all considerably smaller than those inferred for brown dwarf clouds from observations \citep[$\sim$1-5, e.g.][]{stephens2009}, despite physical conditions (e.g. high-g) matching that of brown dwarfs. As Fig. \ref{fig:surfenvar} (and Table \ref{table:all}) shows, this may be due to material properties that control the nucleation rate of particles. The homogeneous nucleation rate (Eq. \ref{eq:homnuc}) is heavily dependent on the value of the exponent in its definition, $-F/kT$, where $F$ is the formation energy of the nucleated critical cluster and $k$ is the Boltzmann constant  -- the greater the absolute value of this term, the lower the rate. Expanding out the exponent in the relevant physical variables, we find that

\begin{equation}
\label{eq:homexpon}
-\frac{F}{kT} =  -\frac{16}{3}\pi \frac{m^2 \sigma_s^3 }{\rho_p^2 (kT)^3 \ln^2{S}}
\end{equation}

\noindent where $m$ is the mass of a condensate molecule and $S$ is the condensate vapor saturation ratio. The strong dependence of the exponent on the surface energy translates to a strong dependence of the cloud distribution on condensate material properties. As $\sigma_s$ increases, the nucleation rate drops precipitously \{and the peak of the nucleation rate shifts to cooler (lower pressure) regions of the atmosphere where higher saturation ratios somewhat offset the increase in $F$ from the increased $\sigma_s$ (Fig. \ref{fig:homnucprofile}). Beyond a certain threshold, increasing $\sigma_s$ no longer results in any significant cloud formation, assuming homogeneous nucleation. The lower nucleation rate with increasing $\sigma_s$ leads to diminutive clouds composed of mostly a small number of large particles and high $f_{\rm sed}$ values. The increase in the size of cloud particles is due to the combined effect of the decrease in the number density of cloud particles and the increase in the condensate vapor density, leading to increased growth rates per cloud particle (Fig. \ref{fig:growparticle}). As $\sigma_s$ can vary by as much as a factor of 10 -- for example between KCl (Eq. \ref{eq:kclsurftens}) and ZnS \citep[1672 ergs cm$^{-2}$;][]{celikkaya1990} -- the homogeneous nucleation rate, and thus the cloud distribution, can vary substantially between different condensates despite similar background atmospheric conditions. 

\subsection{Heterogeneous Nucleation}

\begin{figure}[hbt!]
\centering
\includegraphics[width=0.8 \textwidth]{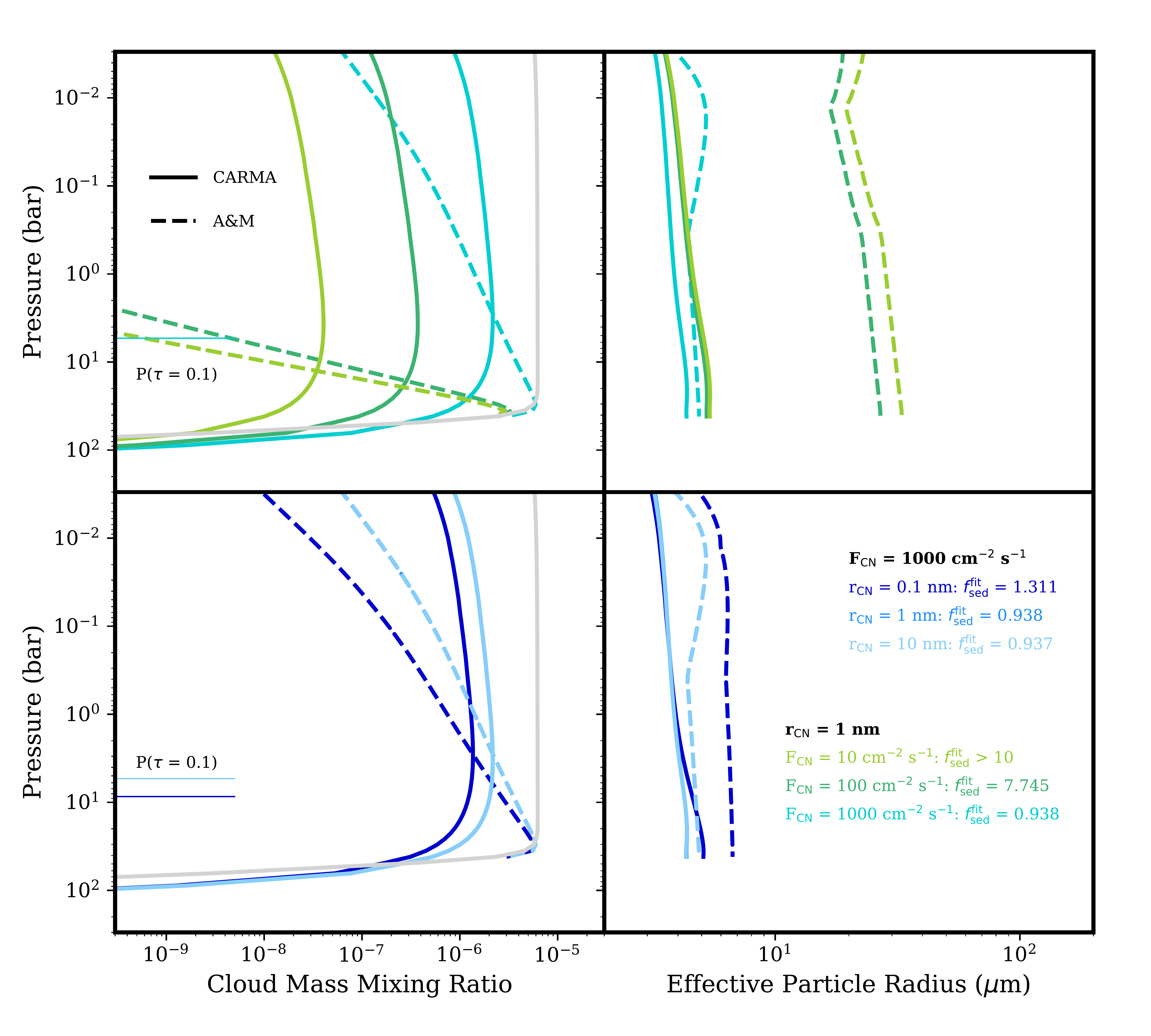}
\caption{Cloud mass mixing ratio $q_c$ (left) and effective cloud particle radius $r_{\rm eff}$ (right) profiles computed by CARMA (solid) compared to that of the best fit $f_{\rm sed}$ AM01 profiles (dashed) as functions of atmospheric pressure level for the heterogeneous nucleation cases where the downward flux of condensation nuclei are varied (top; cases 11-13 in Table \ref{table:all}; downward condensation nuclei flux fixed to 1000 cm$^{-2}$ s$^{-1}$) and where the size of the condensation nuclei are varied (bottom; cases 13-15 in Table \ref{table:all}; condensation nuclei radius fixed to 1 nm). $K_{zz}$ is fixed to $10^8$ cm$^{2}$ s$^{-1}$ and log $g$ is fixed to 5.25. The pressure levels where the cumulative optical depth reaches 0.1 from above are shown as horizontal solid lines with colors corresponding to the $q_c$ and $r_{\rm eff}$ profiles. The 2 lowest condensation nuclei flux clouds never reach a cumulative optical depth of 0.1. The cases where the condensation nuclei radius is 1 nm and 10 nm are nearly identical and so plot on top of each other. }
\label{fig:nucvar}
\end{figure}

We find that the cloud distribution arising from heterogeneous nucleation is dependent on the supply rate and size of condensation nuclei (Fig. \ref{fig:nucvar}; Table \ref{table:all}). The cloud mass density scales with condensation nuclei flux roughly linearly in logspace, though this effect could change at higher condensation nuclei fluxes due to the limited supply of condensate vapor upwelled from depth. The dependence on condensation nuclei radius $r_{CN}$ arises from the Kelvin curvature effect, where the saturation vapor pressure over a curved surface $p_s^{cur}$ is greater than that over a flat surface by an amount defined by

\begin{equation}
\label{eq:kelvin}
\ln \frac{p_s^{cur}}{p_s} = \frac{2M\sigma_s}{\rho_p r_{CN} R T}
\end{equation}

\noindent where $M$ is the molecular weight of the condensate. Thus, smaller condensation nuclei lead to greater saturation vapor pressures, and lower nucleation rates. However, this effect is only relevant at sufficiently small $r$, as the right hand side of Eq. \ref{eq:kelvin} approaches 0 at large $r$. This is reflected in the lower 2 plots of Fig. \ref{fig:nucvar}, where a much greater difference exists between clouds nucleating on 0.1 and 1 nm condensation nuclei than between clouds nucleating on 1 and 10 nm condensation nuclei. 

\subsection{Particle Size Distribution}

\begin{figure}[hbt!]
\centering
\includegraphics[width=0.8 \textwidth]{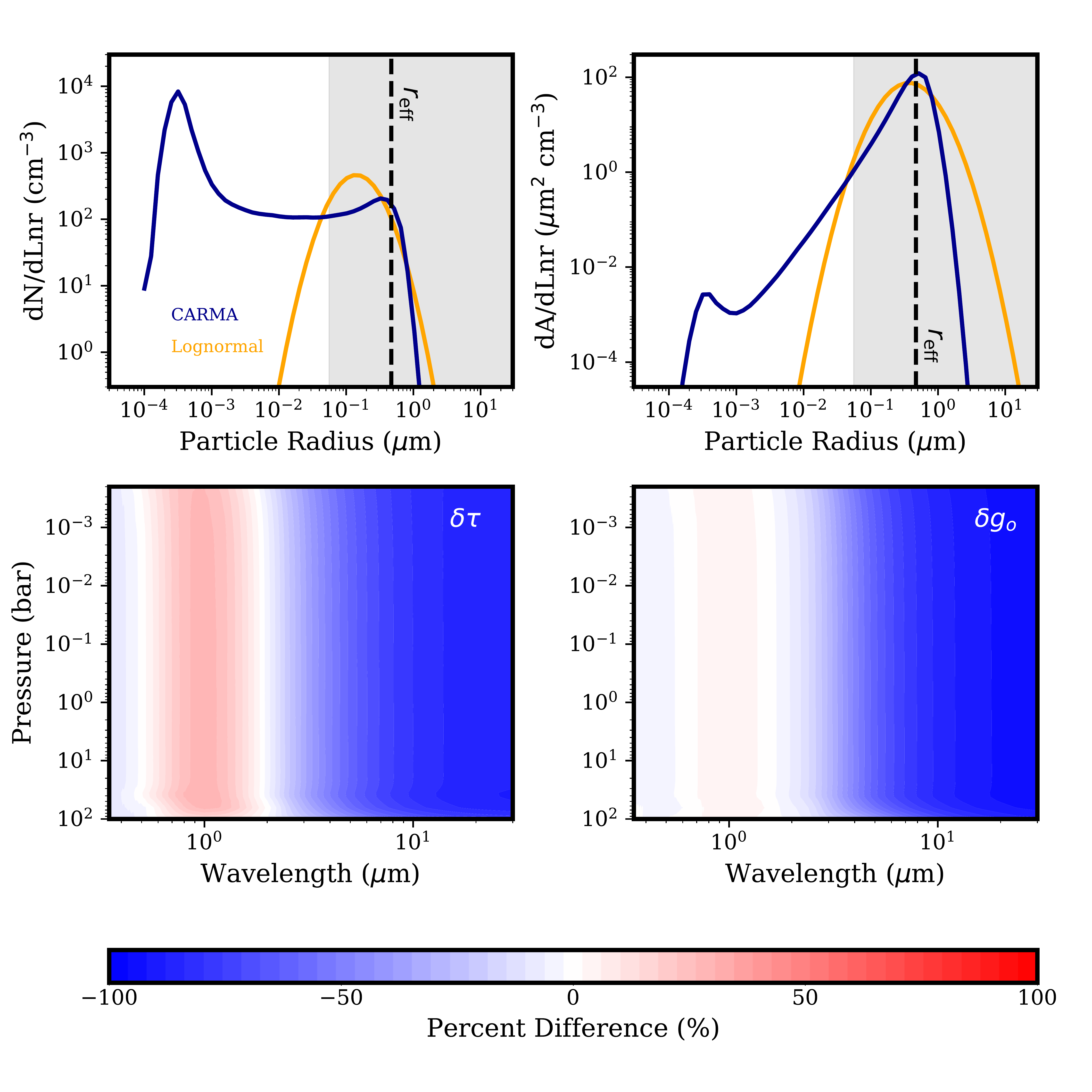}
\caption{(Left, top) Comparison of the binned cloud particle size distribution output of CARMA (blue) and the lognormal parameterization of that distribution (orange) for the $K_{zz}$ = $10^8$ cm$^{2}$ s$^{-1}$ and log $g$ = 5.25 case, at the pressure level where the cumulative optical depth reaches 0.1. The shaded range of radii include $r$ values satisfying $2\pi r / \lambda_{min}$  $>$ $1$, where $\lambda_{min}$ = 0.35 $\mu$m is the lower bound of the domain of the bottom two plots. The $r_{\rm eff}$ value for this pressure level is marked by the vertical dashed line. (Right, top) Same as the left--top plot, but comparing the particle cross sectional area distribution. (Left, bottom) Percent difference in layer optical depth between the binned size distribution and the lognormal parameterization, where a positive difference corresponds to a higher value for the binned distribution. The $r_{\rm eff}$ profile is marked by the dashed line.  (Right, bottom) Same as the left--bottom plot, but for the asymmetry factor.}
\label{fig:sizedist}
\end{figure}

As the particle size distribution is fully resolved in CARMA using mass bins while AM01 relies on a lognormal parameterization of the size distribution, it is useful to assess how much this difference impacts the resulting observable parameters. In order to isolate the effect of the difference in size distribution shape only, we treat only the CARMA results, and compare the binned results with a lognormal distribution computed from them. We also assume that KCl has a constant real refractive index of 1.5, and that it is purely scattering, i.e. its imaginary refractive index is 0 at all wavelengths. The lognormal parameterization of the CARMA results is computed by first calculating $\epsilon q_c$ and $r_{\rm eff}$ at every pressure level according to the procedure given in ${\S}$\ref{sec:modelcomp}, then calculating $r_g$ from $r_{\rm eff}$ following Eq. \ref{eq:reff}. The total number density $N$ is then computed using Eq. 14 from AM01

\begin{equation}
\label{eq:totaln}
N = \frac{3 \epsilon \rho_a q_c}{4 \pi \rho_p r_{g}^3} \exp \left ( -\frac{9}{2} \ln^2 \sigma_g \right )
\end{equation}

\noindent with $\sigma_g$ set to 2 as with AM01. 

The resulting lognormal distribution is drastically different from the bimodal structure of the actual size distribution as computed by CARMA (Fig. \ref{fig:sizedist}), where the lognormal parameterization fails to capture the small particle population originating from nucleation. However, this has minimal impact on the optical properties of the cloud, as the ``nucleation mode'' particles are far too small to significantly affect the oft-observed wavelengths (e.g. visible to thermal-IR). This is shown by the particle cross sectional area distribution in the top right plot of Fig. \ref{fig:sizedist}, which ultimately controls the optical depth of the cloud. Nonetheless, the existence of the nucleation mode shifts the lognormal distribution towards smaller radii compared to the ``condensation mode'' in the CARMA results, which creates differences in optical properties between the binned size distribution and the lognormal parameterization of roughly 100\%, or a factor of 2 at most across all wavelengths, and much smaller differences ($<$20\%) at wavelengths corresponding to particle radii where the optical depth is actually significant ($\leqslant$2 $\mu$m). 

These differences in optical properties are mostly independent of altitude but are dependent on wavelength following the differences in particle number densities and cross sectional areas between the two size distributions as a function of particle size. Particles with radius $r$ notably impact the extinction of light of wavelength $\lambda$ when the shape factor $2\pi r / \lambda$ $\geqslant$ 1, as shown by the shaded region in the top plots of Fig. \ref{fig:sizedist} for $\lambda$ =  0.35 $\mu$m, the lower wavelength bound of the bottom two plots. As such, most of the differences seen in the bottom two plots are due to particles with radii in the shaded region, and the difference with increasing wavelength can be explained by which size distribution dominates as the shaded region moves to larger radii. For example, at 0.35 $\mu$m, the lognormal parameterization slightly dominates over the binned distribution, leading to a negative difference in the optical depth and asymmetry factor. However, as the wavelength is increased such that particles smaller than $\sim$0.4 $\mu$m no longer contribute appreciably to extinction, the higher cross sectional area of the lognormal parameterization versus the binned distribution for particle radii between 0.05 and 0.4 $\mu$m become irrelevant, and the binned distribution dominates, leading to a positive difference. At longer wavelengths, there are significantly more large particles in the lognormal parameterization than the binned distribution, and so the former dominates again, leading to negative differences. No difference exists for the single scattering albedo since we have assumed no absorption. 

\section{Discussion}\label{sec:discussion}

\subsection{Comparison to Previous Works}

As pointed out in ${\S}$\ref{sec:varsurf}, the $f_{\rm sed}$ values associated with brown dwarfs ($\sim$1-5) are considerably higher than the values we have found that best fit our KCl clouds ($\sim$0.1). We inferred then that this may be due to brown dwarf clouds having higher surface energies. The surface energies of high temperature condensates iron and forsterite, for example, are 1720 ergs cm$^{-2}$ \citep{halden1955} and 1280 ergs cm$^{-2}$ \citep{miura2010}, respectively. We can compare the value of the factor $-F/kT$ (Eq. \ref{eq:homexpon}) between KCl and iron and forsterite to quantify the effects of material properties and atmospheric conditions on the nucleation rate. For KCl, we set $\rho_p$ = 1.99 g cm$^{-3}$, $T$ = 850 K, the condensation temperature of KCl at $\sim$1 bar in a solar metallicity atmosphere, $m$ = 74.5$m_p$, where $m_p$ is the proton mass, and use the surface energy expression in Eq. \ref{eq:kclsurftens}. For iron and forsterite, we set $\rho_p$ to 7.875 and 3.27 g cm$^{-3}$, respectively, $T$ = 1800 K, roughly the condensation temperature of iron and forsterite at $\sim$1 bar, and $m$ to 55.845$m_p$ and 140.69$m_p$, respectively. The resulting $-F/kT$ values for iron and forsterite are factors of 19 and 284 higher than that of KCl, which greatly decrease the homogeneous nucleation rates of these two species, as $-F/kT$ is in the exponential. Iron and forsterite clouds could also form via heterogeneous nucleation on condensation nuclei composed of TiO$_2$, as has been proposed by e.g. \citet{helling2006}. 

\citet{morley2012} showed that Na$_{2}$S clouds forming at 600 K with $f_{\rm sed}$ $\sim$ 4-5 could explain the color of T dwarfs, indicating that Na$_{2}$S's surface energy (currently unknown) may also be large, especially if it is similar to ZnS. Note that \citet{morley2012} also included KCl in their model, but the high $f_{\rm sed}$ rendered their KCl clouds optically thin. Comparing the KCl clouds between their model and our results is also difficult as the $K_{zz}$ profile they used is not given. In contrast, \citet{morley2013,morley2015} found that KCl and ZnS clouds in the atmosphere of GJ 1214 b must have low $f_{\rm sed}$ values ($\sim$0.1-0.01) to fit the observed transmission spectrum, which is consistent with our results. However, those models had higher metallicities, which may not be completely compatible with the results of our solar metallicity simulations. It remains to be seen what the effect of having low $f_{\rm sed}$ for KCl -- or different $f_{\rm sed}$ for different clouds -- would be on brown dwarf colors. A thicker cloud would lead to redder colors, but the smaller particles could result in the clouds becoming optically thin in the NIR, making the objects appear bluer. 

\subsection{Application to the Brown Dwarf L/T Transition}

A particular application of our results is on the brown dwarf L/T transition, where the near-IR colors of brown dwarfs switch from a reddening trend with decreasing luminosity to a trend towards the blue \citep[see reviews by][]{kirkpatrick2005,helling2014review}. \citet{saumon2008} and \citet{stephens2009} found that the AM01 model can reproduce the L/T transition if the $f_{\rm sed}$ values of clouds therein increased from $\sim$2 to $\geqslant$4-5. Alternatively, \citet{marley2010} showed that the L/T transition can be caused by the appearance of holes in the clouds, with the $f_{\rm sed}$ values of the remaining (patchy) clouds staying $\sim$2. As the silicate and iron clouds sink to higher pressure levels in the atmosphere, the increase in local temperature should lead to an increase in the nucleation rate and thus smaller particles. However, this effect may be countered by the increased abundance of condensable material at depth, thereby growing particles at a faster rate, leading to larger particles and higher $f_{\rm sed}$ values. This latter process may be more dominant, as the increase in condensation temperature is small compared to the exponential increase in atmospheric density with increasing depth. \{To test how $f_{\rm sed}$ evolves with decreasing $T_{\rm eff}$ for a single cloud species, we examine the behavior of corundum (Al$_{2}$O$_{3}$), which is an important condensate in L dwarf atmospheres. For our singular purpose of examining trends in $f_{\rm sed}$, we assume particle formation via homogeneous nucleation, as we have done for KCl. However, for future, more detailed exploration of corundum cloud formation we note that equilibrium chemistry arguments suggest that other aluminum-bearing species, such as Al$_{2}$O$_{2}$ and AlOH may be more abundant in the gas phase, limiting corundum cloud formation by direct phase change \citep{helling2013}.

For the saturation vapor pressure of corundum we use 

\begin{equation}
\label{eq:al2o3svp}
\log{p_s^{\rm Al_{2}O_{3}} (\rm bars)} = 17.7 - 45892.6/T(\rm K)
\end{equation}

\noindent which is derived from the condensation curve of corundum given in \citet{wakeford2017} and the solar abundance of Al from \citet{lodders2010}. We use a surface energy of 900 ergs cm$^{-2}$ \citep{dobrovinskaya2009}. The model atmospheres used are from the same grid as those used in the other simulations in this work, with log $g$ = 4.75. We assume a constant $K_{zz}$ value of $10^{8}$ cm$^{2}$ s$^{-1}$ for all cases. As $T_{\rm eff}$ is decreased, the cloud base sinks deeper into the atmosphere, the particles at the cloud base become larger from the increased supply of condensate vapor, and the best-fit $f_{\rm sed}$ increases. The more massive iron and silicate clouds behaving similarly could thus explain the increase in $f_{\rm sed}$ with stellar type at the L/T transition. 

\begin{figure}[hbt!]
\centering
\includegraphics[width=0.8 \textwidth]{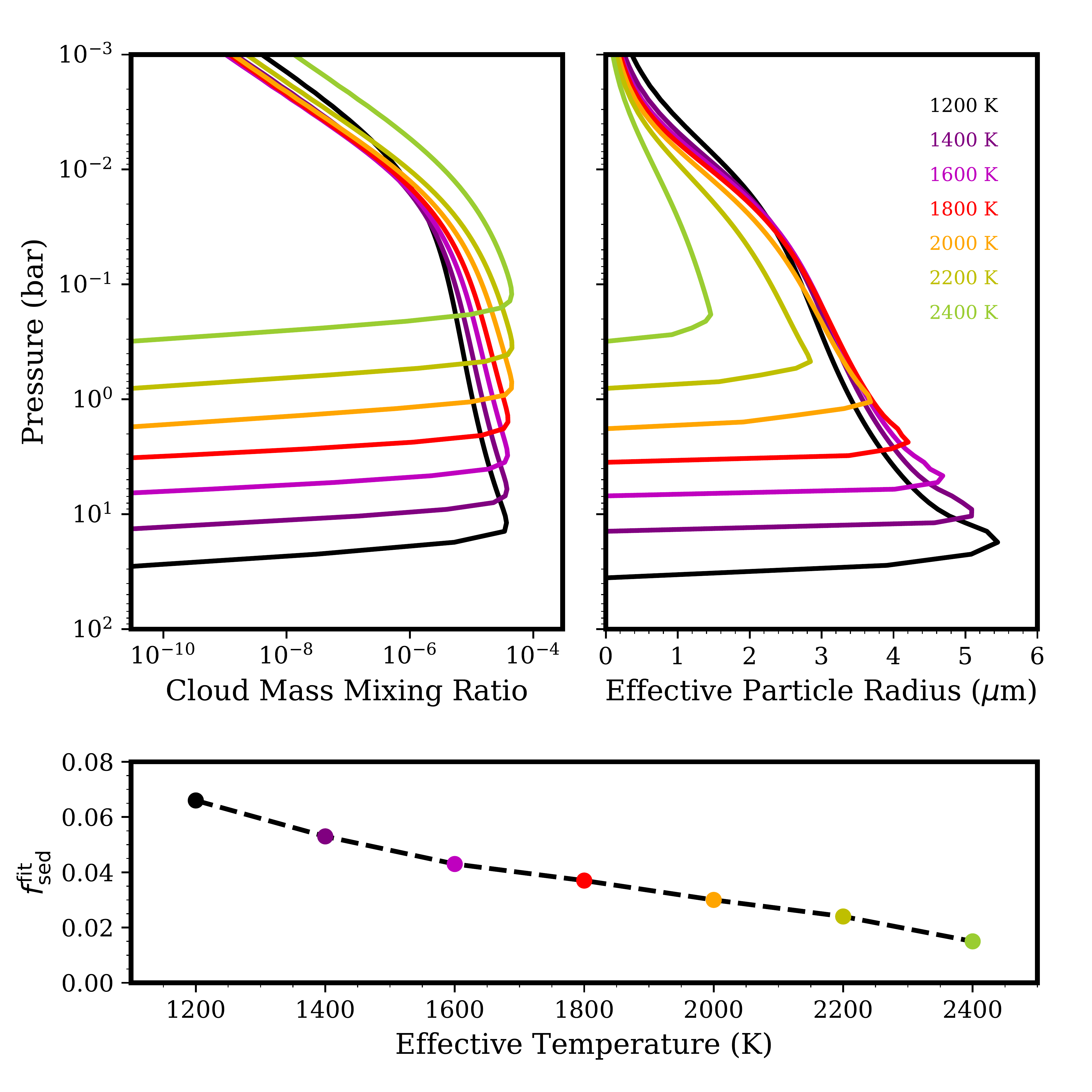}
\caption{Cloud mass mixing ratio $q_c$ (left, top) and effective cloud particle radius $r_{\rm eff}$ (right, top) profiles computed by CARMA for homogeneous nucleation of corundum clouds over a range of effective temperatures, along with the best-fit $f_{\rm sed}$ of each simulated cloud distribution (bottom). We assume log $g$ = 4.75 and $K_{zz}$ = $10^{8}$ cm$^{2}$ s$^{-1}$.}
\label{fig:tempevol}
\end{figure}

Alternatively, if silicate and iron clouds depend on condensation nuclei to form, then the L/T transition could be explained by their disappearance, such as due to cold-trapping of the material making up the condensation nuclei \citep[e.g. TiO$_2$;][]{lee2015b} at depth with decreasing effective temperature, essentially ``starving'' the clouds. 

The reliance on condensation nuclei could explain another anomalous observation: the extreme reddening of young, directly imaged giant exoplanets, which has been attributed to persistent cloudiness even at low luminosity \citep[see for example Fig. 7 in][]{bowler2016}. While low gravity objects are intrinsically hotter than high gravity objects at the same effective temperature, the cloudiness can perhaps be enhanced by a high flux of interplanetary dust from their young stellar systems that create large volumes of meteoritic smoke particles within their atmospheres, which can serve as condensation nuclei. If this is the case, then young substellar objects located in young, dusty stellar systems, should be cloudier than their isolated counterparts, regardless of mass. We will investigate this mechanism in a future publication. 

\section{Conclusions}\label{sec:conclusions}


\{To the extent that the CARMA results represent physical cloud distributions in the limit of horizontally homogeneous clouds, we find that the $f_{\rm sed}$ which best describes any given profile is most sensitive to the material properties and formation pathways of the cloud itself. In particular, the strong dependence of the cloud particle nucleation rate on the cloud material's surface energy requires improved constraints on their values via laboratory measurements and \textit{ab initio} calculations, as surface energies can vary by more than an order of magnitude between different materials leading to several orders of magnitude differences in cloud opacity. Meanwhile, the dependence of cloud opacity on the source flux and size of condensation nuclei necessitates theoretical studies of these quantities in exoplanet and brown dwarf atmospheres across a range of environments. For example, \citet{lee2018} investigated whether certain cloud species, including KCl can act as condensation nuclei for other cloud species due to their relatively high nucleation rates at low supersaturations. \citet{lavvas2017} discussed the possibility of particles recondensed from meteoroid ablation and soot aerosols arising from photochemistry acting as nucleation sites for silicate clouds in hot Jupiter atmospheres. Extension of this latter work to giant planets in young, dusty systems could yield significant insights into the physical processes governing their near-IR colors. 

\{We further find that $f_{\rm sed}$ is dependent on the strength of vertical mixing, as parameterized by $K_{zz}$, but to a lesser extent than on material properties. $f_{\rm sed}$ is only slightly dependent on the local gravitational acceleration given constant $K_{zz}$ values, though a more realistic representation of $K_{zz}$ that scales with gravity (e.g. the $K_{zz}$ parameterization in AM01) may yield different conclusions. Thus, improved coupling between gravitational acceleration and atmospheric dynamics is needed to better understand cloud distributions in 1D atmospheric models. On the whole, we find that AM01 can roughly reproduce the spatial distribution and mean size of cloud particles in substellar atmospheres computed with consideration of more detailed cloud microphysics.

\{The trends seen in $f_{\rm sed}$ with material properties, formation pathway, strength of mixing, and gravity can lead to the development of improved cloud models that take into account microphysical processes such as nucleation and condensation, but which are still relatively simple enough to run rapidly. These models could improve our understanding of exoplanet cloudiness, the brown dwarf L/T transition, and the extreme redness of young substellar objects by providing a physical basis for values of $f_{\rm sed}$ and how it varies with planetary parameters. For example, we predict that $f_{\rm sed}$ should generally increase with increasing depth of the cloud base in an atmosphere (i.e. decreasing $T_{\rm eff}$) for the same cloud species due to higher abundances of condensate vapor at depth aiding growth of larger cloud particles, in line with $f_{\rm sed}$ trends inferred from observations. This process is not captured in simpler models that do not take into account nucleation and condensation, and could be a pathway towards understanding the L/T transition. Future models will also need to refine treatments of the spatial inhomogeneity of clouds and particle distributions \citep{marley2010,parmentier2016,line2016,macdonald2017}, which is not well captured in 1D models. These new pseudo-3D models will need to parameterize the 3D temperature structure of substellar objects and take into account the horizontal transport time scales in their atmospheres.

\acknowledgments

We thank C. V. Morley and V. Parmentier for enlightening discussions. P. Gao acknowledges funding support from the NASA Postdoctoral Program and the 51 Pegasi b Fellowship in Planetary Astronomy from the Heising-Simons Foundation.  M. S. Marley acknowledges the support of the NASA Exoplanet Research and Astrophysics Theory Programs.

\appendix
\section{CARMA Overview}\label{sec:carma}

\subsection{Nucleation}

Classical theories of homogeneous and heterogeneous nucleation are used for computing the rates of new particle generation \citep{pruppacher1978,lavvas2011}. For homogeneous nucleation, the rate $J_{hom}$ in units of new particles per volume per unit time is given as

\begin{equation}\label{eq:homnuc}
J_{hom} = 4\pi a_c^2 \Phi Z n \exp{(-F/kT)},
\end{equation}

\noindent where $n$ is the number density of condensate vapor molecules, $k$ is the Boltzmann constant, and $T$ is the temperature. $a_c$ is the critical particle radius defined as 

\begin{equation}
a_c = \frac{2M\sigma_s}{\rho_p R T \ln{S}},
\end{equation}

\noindent where $M$, $\sigma_s$, $\rho_p$, and $S$ are the molecular weight, surface tension (energy), mass density, and saturation ratio of the condensate, respectively, and $R$ is the universal gas constant. $F$ is the energy of formation of a particle with radius $a_c$, given by

\begin{equation}
F = \frac{4}{3}\pi \sigma_s a_c^2.
\end{equation}

\noindent In other words, classical homogeneous nucleation theory relates the rate of particle formation with the energy of formation. The greater the energy (i.e. the lower the supersaturation), the lower the nucleation rate. The energy of formation is defined as the balance between the increase in energy associated with enlarging a particle's surface to add vapor molecules and the decrease in energy due to the increase in the volume of the particle. Particles with a radius of $a_c$ is at the cusp of this balance, where further growth leads to a net decrease in energy of formation, and therefore continued existence. $\Phi$ is the diffusion rate of vapor molecules to the forming particle, defined as

\begin{equation}
\Phi = \frac{p}{\sqrt{2\pi m k T}},
\end{equation}

\noindent where $p$ is the atmospheric pressure and $m$ is the mass of a vapor molecule. Finally, $Z$ is the Zeldovich factor, which takes into account non-equilibrium effects, such as the evaporation of just--formed particles occurring at the same time as particle formation. $Z$ is given by

\begin{equation}\label{eq:zeld}
Z = \sqrt{\frac{F}{3 \pi k T g_m^2}},
\end{equation}

\noindent where $g_m$ is the number of molecules in particles with radius $a_c$. 

The rate of heterogeneous nucleation, $J_{het}$ is defined as,

\begin{equation}\label{eq:hetnuceq}
J_{het} = 4\pi^2 r_{CN}^2 a_c^2 \Phi c_{surf} Z \exp{(-Ff/kT)},
\end{equation}

\noindent where $a_c$ is the critical radius of the initial cluster of condensate molecules (``germ'') on the nucleating surface, $r_{CN}$ is the radius of the condensation nuclei, $f$ is a shape factor given by

\begin{equation}
2f = 1 + \left ( \frac{1 - \mu x}{\phi} \right )^3 + x^3 \left (2 - 3 f_0 + f_0^3 \right ) + 3 \mu x^2 \left (f_0 - 1 \right ),
\end{equation}

\noindent where $\mu$ is the cosine of the contact angle between the condensate and the nucleating surface, and 

\begin{equation}
x = r/a_c
\end{equation}

\begin{equation}
\phi = \sqrt{1 -  2\mu x + x^2}
\end{equation}

\begin{equation}
f_0 = (x - \mu)/\phi
\end{equation}

\noindent and $c_{surf}$ is the number density of condensate molecules on the nucleating surface, given by

\begin{equation}
c_{surf} = \frac{\Phi}{\nu}\exp{(F_{des}/kT)},
\end{equation}

\noindent where $\nu$ is the oscillation frequency of adsorbed vapor molecules on the nucleating surface and $F_{des}$ is the desorption energy of that molecule. The units of $J_{het}$ is critical germs per condensation nucleus, and thus to get the number of newly nucleated particles per volume per unit time $J_{het}$ must be multiplied by the number density of condensation nuclei. 

\subsection{Condensation \& Evaporation}

Condensational growth and evaporation in CARMA takes into account the diffusion of condensate molecules to and away from the cloud particle, such that the rate of change in mass $m_p$ of the particle is 

\begin{equation}\label{eq:cond1}
\frac{dm_p}{dt} = 4 \pi R_d^2 D \frac{d\rho_v}{dR_d},
\end{equation}

\noindent where $R_d$ is the distance away from the center of the particle, $D$ is the molecular diffusion coefficient of the condensate vapor through the background atmosphere, and $\rho_v$ is the mass density of the condensate vapor. Addition of molecules to a particle releases latent heat, and vice versa for the removal of molecules. The rate of change in temperature of the particle, $T_p$ arising from condensation and evaporation is

\begin{equation}\label{eq:cond2}
m_p C_p \frac{dT_p}{dt} = L\frac{dm_p}{dt} - \frac{dQ}{dt},
\end{equation}

\noindent where $C_p$ is the heat capacity of the particle, $L$ is the latent heat of evaporation of the condensate, and $dQ/dt$ is the conductive cooling rate of the particle given by

\begin{equation}\label{eq:cond3}
\frac{dQ}{dt} = - 4 \pi R_d^2 \kappa_a \frac{dT}{dR_d},
\end{equation}

\noindent where $\kappa_a$ is the thermal conductivity of the atmosphere. Combining Eqs. \ref{eq:cond1}--\ref{eq:cond3}, along with the ideal gas law and the Clausius--Clapeyron equation, 

\begin{equation}\label{eq:dmdt}
\frac{dm_p}{dt} = \frac{4 \pi r  D p_s (S - 1)}{\frac{RT}{M} + \frac{D L p_s}{\kappa_a T} \left ( \frac{L M}{R T} - 1 \right )},
\end{equation}

\noindent where $p_{sat}$ is the saturation vapor pressure of the condensate \citep[see][for a full derivation]{jacobson2005}. CARMA makes the assumption that $LM/RT - 1$ $\sim$ $LM/RT$, and takes into account several additional effects, such that Eq. \ref{eq:dmdt} becomes 

\begin{equation}
\frac{dm_p}{dt} = \frac{4 \pi r  D' p_s  (S - A_k)}{\frac{RT}{MF_v} + \frac{D' M L^2 p_s}{\kappa_a' RT^2 F_t} },
\end{equation}

\noindent where $A_k$ is the Kelvin factor, which takes into account the curvature of the particle surface, and is given by

\begin{equation}
A_k = \exp{\left ( \frac{2M\sigma_s}{\rho_p R T r} \right )}.
\end{equation}

\noindent $D'$ and $\kappa_a'$ are the molecular diffusional coefficient of the condensate vapor through the atmosphere and the thermal conductivity of the atmosphere, respectively, modified to account for gas kinetics effects near the surface of the particle, and are expressed as

\begin{equation}
D' = \frac{D}{1 + \lambda Kn^c}
\end{equation}
\begin{equation}
\kappa_a' = \frac{\kappa_a}{1 + \lambda_t Kn^c_t},
\end{equation}

\noindent where $\lambda$ and $\lambda_t$ are given by

\begin{equation}
\lambda = \frac{1.33Kn^c+ 0.71}{Kn^c+1} + \frac{4(1-\alpha_s)}{3\alpha_s}
\end{equation}
\begin{equation}
\lambda_t = \frac{1.33Kn^c_t+ 0.71}{Kn^c_t+1} + \frac{4(1-\alpha_t)}{3\alpha_t},
\end{equation}

\noindent where $\alpha_s$ and $\alpha_t$ are the sticking coefficient and the thermal accommodation coefficient, respectively, which are set to 1, and $Kn^c$ and $Kn^c_t$ are Knudsen numbers of the condensing gas with respect to the particle, defined as 

\begin{equation}\label{eq:knc}
Kn^c = \frac{3 D}{r} \sqrt{\frac{\pi M}{8RT}} 
\end{equation}
\begin{equation}\label{eq:knct}
Kn^c_t = \frac{Kn^c \kappa_a}{rD \rho_a \left (C_p - \frac{R}{2\mu_a} \right )},
\end{equation}

\noindent where $\rho_a$ is the atmospheric mass density and $\mu_a$ is the mean molecular weight of the atmosphere. Finally, $F_v$ and $F_t$ are the ventilation factors that account for the air density variations around a particle as it sediments in an atmosphere, which increases density in front of it and lowers the density behind it \citep{toon1989,lavvas2011}. 

CARMA uses the piecewise parabolic method for treating advection of particles \citep{colella1984} to apply the rate of change in mass $dm_p/dt$ to evolving the binned particle distribution. In other words, CARMA treats the fluxes between mass bins as if they were fluxes between altitude levels. 

\subsection{Coagulation}

Coagulation of particles occurs when the particles bump into each other due to Brownian motion and stick. The rate of coagulation of particles from two populations is given by the product of the coagulation kernel, $K_b$, the number density of one of the groups of particles, and the number density of the other group of particles. Given particle populations 1 and 2, the coagulation kernel is defined by

\begin{equation}
K_b = \frac{4 \pi (D_1 + D_2)(r_1 + r_2)}{\frac{r_1 + r_2}{r_1 + r_2 + \sqrt{\delta^2_1 + \delta^2_2}} + \frac{4 (D_1 + D_2)}{(r_1 + r_2)\sqrt{V^2_1 + V^2_2}}},
\end{equation}

\noindent where $D_x$, $r_x$, $V_x$, and $\delta_x$, $x$ = 1,2 are the molecular diffusion coefficients, radii, thermal velocities, and interpolation factors of particles from populations 1 and 2, respectively. $D_x$ is defined as 

\begin{equation}
D_x = \frac{kT\beta}{6 \pi \eta r_x},
\end{equation}

\noindent where $\beta$ is the Cunningham slip correction factor (see below) and $\eta$ is the dynamic viscosity. $V_x$ is given by

\begin{equation}
V_x = \sqrt{\frac{8kT}{\pi m_{px}}},
\end{equation}

\noindent where $m_{px}$, $x$ = 1,2 is the mass of the particle. Finally, $\delta_x$ is expressed as

\begin{equation}
\delta_x = \frac{(2r_x + \lambda_x)^3 - (4r^2_x + \lambda^2_x)^{3/2}}{6 r_x \lambda_x} - 2 r_x,
\end{equation}

\noindent where $\lambda_x$ is the particle mean free path and is given by

\begin{equation}
\lambda_x = \frac{8 D_x}{\pi V_x}.
\end{equation}

\noindent $\delta_x$ interpolates between the continuum and kinetic regimes \citep{lavvas2010}.

\subsection{Vertical Transport}\label{sec:carmavt}

The sedimentation velocity $v_f$ of a spherical particle of radius $r$ is given by Stoke's fall velocity,

\begin{equation}
v_f = \frac{2}{9} \frac{\rho_p g r^2 \beta}{\eta} ,
\end{equation}

\noindent where $g$ is the gravitational acceleration and $\beta$ again is the Cunningham slip correction factor given by 

\begin{equation}
\beta = 1 + 1.246 Kn + 0.42 Kn e^{-0.87/Kn},                                      
\end{equation}

\noindent where $Kn$ is the Knudsen number of the particle defined as the ratio of the atmospheric mean free path $l$ to $r$, where $l$ can be written as

\begin{equation}
l = \frac{2\eta}{\rho_a}\sqrt{\frac{\pi \mu_a}{8 R T}}.
\end{equation}                                                             

\noindent Note that $Kn$ is different from $Kn^c$ and $Kn^c_t$ defined in Eqs. \ref{eq:knc}--\ref{eq:knct}. $\beta$ $\sim$ 1 in the continuum regime ($Kn$ $\ll$ 1). In the kinetics regime, $\beta$ is large and mostly linear with $Kn$. Taking this into account, the sedimentation velocity becomes

\begin{equation}
v = A \frac{\rho_p g r}{\rho_a}\sqrt{\frac{\pi \mu_a}{2 R T}} ,                                                      
\end{equation}

\noindent where A is a constant that is $\sim$0.5.

The velocity associated with advection is calculated from a user defined wind speed via the piecewise parabolic method \citep{colella1984}. Diffusive velocities are assumed to be dominated by eddy diffusion, and are calculated and combined with advective velocities to solve for new particle and gas distributions using the algorithms outlined in \citet{toon1988}. The eddy diffusion coefficient as a function of altitude is defined by the user. 



\vspace{5mm}

\software{CARMA, eddysed}

\bibliography{references}

\begin{thebibliography}{}
\expandafter\ifx\csname natexlab\endcsname\relax\def\natexlab#1{#1}\fi

\bibitem[{Ackerman {et~al.}(1995)Ackerman, Hobbs, \& Toon}]{ackerman1995}
Ackerman, A.~S., Hobbs, P.~V., \& Toon, O.~B. 1995, Journal of the Atmospheric
  Sciences, 52, 1204

\bibitem[{Ackerman \& Marley(2001)}]{ackerman2001}
Ackerman, A.~S., \& Marley, M.~S. 2001, The Astrophysical Journal, 556, 872

\bibitem[{Ackerman {et~al.}(1993)Ackerman, Toon, \& Hobbs}]{ackerman1993}
Ackerman, A.~S., Toon, O.~B., \& Hobbs, P.~V. 1993, Journal of the Atmospheric
  Sciences, 52, 1204

\bibitem[{{Allard} {et~al.}(2001){Allard}, {Hauschildt}, {Alexander},
  {Tamanai}, \& {Schweitzer}}]{allard2001}
{Allard}, F., {Hauschildt}, P.~H., {Alexander}, D.~R., {Tamanai}, A., \&
  {Schweitzer}, A. 2001, \apj, 556, 357

\bibitem[{Bardeen {et~al.}(2008)Bardeen, Toon, Jensen, Marsh, \&
  Harvey}]{bardeen2008}
Bardeen, C.~G., Toon, O.~B., Jensen, E.~J., Marsh, D.~R., \& Harvey, V.~L.
  2008, Journal of Geophysical Research, 113, D17202

\bibitem[{{Benneke}(2015)}]{benneke2015}
{Benneke}, B. 2015, ArXiv e-prints, arXiv:1504.07655

\bibitem[{{Biller}(2017)}]{biller2017}
{Biller}, B. 2017, The Astronomical Review, 13, 1

\bibitem[{B\l{}aszczyszyn {et~al.}(1995)B\l{}aszczyszyn, B\l{}aszczyszynowa, \&
  Gubernator}]{blaszczyszyn1995}
B\l{}aszczyszyn, R., B\l{}aszczyszynowa, M., \& Gubernator, W. 1995, Acta
  Physica Polonica A, 88, 1151

\bibitem[{{Bowler}(2016)}]{bowler2016}
{Bowler}, B.~P. 2016, \pasp, 128, 102001

\bibitem[{{Buenzli} {et~al.}(2012){Buenzli}, {Apai}, {Morley}, {Flateau},
  {Showman}, {Burrows}, {Marley}, {Lewis}, \& {Reid}}]{buenzli2012}
{Buenzli}, E., {Apai}, D., {Morley}, C.~V., {et~al.} 2012, \apjl, 760, L31

\bibitem[{{Burgasser} {et~al.}(2011){Burgasser}, {Cushing}, {Kirkpatrick},
  {Gelino}, {Griffith}, {Looper}, {Tinney}, {Simcoe}, {Bochanski}, {Skrutskie},
  {Mainzer}, {Thompson}, {Marsh}, {Bauer}, \& {Wright}}]{burgasser2011}
{Burgasser}, A.~J., {Cushing}, M.~C., {Kirkpatrick}, J.~D., {et~al.} 2011,
  \apj, 735, 116

\bibitem[{Celikkaya \& Akinc(1990)}]{celikkaya1990}
Celikkaya, A., \& Akinc, M. 1990, Journal of the American Ceramic Society, 73,
  2360

\bibitem[{{Colaprete} {et~al.}(1999){Colaprete}, {Toon}, \&
  {Magalh{\~a}es}}]{colaprete1999}
{Colaprete}, A., {Toon}, O.~B., \& {Magalh{\~a}es}, J.~A. 1999, Journal of
  Geophysical Research, 104, 9043

\bibitem[{Colella \& Woodward(1984)}]{colella1984}
Colella, P., \& Woodward, P.~R. 1984, Journal of Computational Physics, 54, 174

\bibitem[{{Cushing} {et~al.}(2010){Cushing}, {Saumon}, \&
  {Marley}}]{cushing2010}
{Cushing}, M.~C., {Saumon}, D., \& {Marley}, M.~S. 2010, \aj, 140, 1428

\bibitem[{{Demory} {et~al.}(2013){Demory}, {de Wit}, {Lewis}, {Fortney},
  {Zsom}, {Seager}, {Knutson}, {Heng}, {Madhusudhan}, {Gillon}, {Barclay},
  {Desert}, {Parmentier}, \& {Cowan}}]{demory2013}
{Demory}, B.-O., {de Wit}, J., {Lewis}, N., {et~al.} 2013, \apjl, 776, L25

\bibitem[{Dobrovinskaya {et~al.}(2009)Dobrovinskaya, Lytvynov, \&
  Pishchik}]{dobrovinskaya2009}
Dobrovinskaya, E.~R., Lytvynov, L.~A., \& Pishchik, V. 2009, Properties of
  Sapphire (Boston, MA: Springer US), 55--176

\bibitem[{{Esplin} {et~al.}(2016){Esplin}, {Luhman}, {Cushing},
  {Hardegree-Ullman}, {Trucks}, {Burgasser}, \& {Schneider}}]{esplin2016}
{Esplin}, T.~L., {Luhman}, K.~L., {Cushing}, M.~C., {et~al.} 2016, \apj, 832,
  58

\bibitem[{Ethington(1990)}]{ethington1990}
Ethington, E.~F. 1990, Interfacial contact angle measurements of water,
  mercury, and 20 organic liquids on quartz, calcite, biotite, and
  Ca-montmorillonite substrates, Tech. rep., report

\bibitem[{{Fortney} {et~al.}(2006){Fortney}, {Saumon}, {Marley}, {Lodders}, \&
  {Freedman}}]{fortney2006}
{Fortney}, J.~J., {Saumon}, D., {Marley}, M.~S., {Lodders}, K., \& {Freedman},
  R.~S. 2006, \apj, 642, 495

\bibitem[{Gao {et~al.}(2014)Gao, Zhang, Crisp, Bardeen, \& Yung}]{gao2014}
Gao, P., Zhang, X., Crisp, D., Bardeen, C.~G., \& Yung, Y.~L. 2014, Icarus,
  231, 83

\bibitem[{{Gao} {et~al.}(2017){Gao}, {Fan}, {Wong}, {Liang}, {Shia}, {Kammer},
  {Yung}, {Summers}, {Gladstone}, {Young}, {Olkin}, {Ennico}, {Weaver}, \&
  {Stern}}]{gao2017pluto}
{Gao}, P., {Fan}, S., {Wong}, M.~L., {et~al.} 2017, \icarus, 287, 116

\bibitem[{{Gelino} {et~al.}(2014){Gelino}, {Smart}, {Marocco}, {Kirkpatrick},
  {Cushing}, {Mace}, {Mendez}, {Tinney}, \& {Jones}}]{gelino2014}
{Gelino}, C.~R., {Smart}, R.~L., {Marocco}, F., {et~al.} 2014, \aj, 148, 6

\bibitem[{Halden \& Kingery(1955)}]{halden1955}
Halden, F.~A., \& Kingery, W.~D. 1955, The Journal of Physical Chemistry, 59,
  557

\bibitem[{{Heinze} {et~al.}(2013){Heinze}, {Metchev}, {Apai}, {Flateau},
  {Kurtev}, {Marley}, {Radigan}, {Burgasser}, {Artigau}, \&
  {Plavchan}}]{heinze2013}
{Heinze}, A.~N., {Metchev}, S., {Apai}, D., {et~al.} 2013, \apj, 767, 173

\bibitem[{{Helling} \& {Casewell}(2014)}]{helling2014review}
{Helling}, C., \& {Casewell}, S. 2014, \aapr, 22, 80

\bibitem[{{Helling} {et~al.}(2008{\natexlab{a}}){Helling}, {Dehn}, {Woitke}, \&
  {Hauschildt}}]{helling2008c}
{Helling}, C., {Dehn}, M., {Woitke}, P., \& {Hauschildt}, P.~H.
  2008{\natexlab{a}}, \apjl, 675, L105

\bibitem[{{Helling} \& {Fomins}(2013)}]{helling2013}
{Helling}, C., \& {Fomins}, A. 2013, Philosophical Transactions of the Royal
  Society of London Series A, 371, 20110581

\bibitem[{{Helling} \& {Woitke}(2006)}]{helling2006}
{Helling}, C., \& {Woitke}, P. 2006, Astronomy \& Astrophysics, 455, 325

\bibitem[{{Helling} {et~al.}(2008{\natexlab{b}}){Helling}, {Ackerman},
  {Allard}, {Dehn}, {Hauschildt}, {Homeier}, {Lodders}, {Marley}, {Rietmeijer},
  {Tsuji}, \& {Woitke}}]{helling2008a}
{Helling}, C., {Ackerman}, A., {Allard}, F., {et~al.} 2008{\natexlab{b}},
  Monthly Notices of the Royal Astronomical Society, 391, 1854

\bibitem[{{Henderson} {et~al.}(2017){Henderson}, {Skemer}, {Morley}, \&
  {Fortney}}]{henderson2017}
{Henderson}, C.~S., {Skemer}, A.~J., {Morley}, C.~V., \& {Fortney}, J.~J. 2017,
  \mnras, 470, 4557

\bibitem[{Jacobson(2005)}]{jacobson2005}
Jacobson, M.~Z. 2005, Fundmentals of Atmospheric Modeling (Cambridge University
  Press, Cambridge, UK)

\bibitem[{James {et~al.}(1997)James, Toon, \& Schubert}]{james1997}
James, E.~P., Toon, O.~B., \& Schubert, G. 1997, Icarus, 129, 147

\bibitem[{Janz \& Dijkhuis(1969)}]{janz1969}
Janz, G.~J., \& Dijkhuis, C. G.~M. 1969, NSRDS-NBS 28 Report, Molten Salts:
  Volume 2 (Natl. Bur. Standards, Washington, D. C., USA.)

\bibitem[{Jensen {et~al.}(1994)Jensen, Toon, Westphal, Kinne, \&
  Heymsfield}]{jensen1994carma}
Jensen, E.~J., Toon, O.~B., Westphal, D.~L., Kinne, S., \& Heymsfield, A.~J.
  1994, Journal of Geophysical Research, 99, 10421

\bibitem[{{Kirkpatrick}(2005)}]{kirkpatrick2005}
{Kirkpatrick}, J.~D. 2005, \araa, 43, 195

\bibitem[{Kreidberg {et~al.}(2014)Kreidberg, Bean, D{\'{e}}sert, Benneke,
  Deming, Stevenson, Seager, Berta-Thompson, Seifahrt, \&
  Homeier}]{kreidberg2014}
Kreidberg, L., Bean, J.~L., D{\'{e}}sert, J.-M., {et~al.} 2014, Nature, 505, 69

\bibitem[{Lavvas {et~al.}(2011)Lavvas, Griffith, \& Yelle}]{lavvas2011}
Lavvas, P., Griffith, C.~A., \& Yelle, R.~V. 2011, Icarus, 215, 732

\bibitem[{Lavvas \& Koskinen(2017)}]{lavvas2017}
Lavvas, P., \& Koskinen, T. 2017, The Astrophysical Journal, 847, 32

\bibitem[{Lavvas {et~al.}(2010)Lavvas, Yelle, \& Griffith}]{lavvas2010}
Lavvas, P., Yelle, R.~V., \& Griffith, C.~A. 2010, Icarus, 210, 832

\bibitem[{{Lee} {et~al.}(2016){Lee}, {Dobbs-Dixon}, {Helling}, {Bognar}, \&
  {Woitke}}]{lee2016}
{Lee}, G., {Dobbs-Dixon}, I., {Helling}, C., {Bognar}, K., \& {Woitke}, P.
  2016, \aap, 594, A48

\bibitem[{{Lee} {et~al.}(2015){Lee}, {Helling}, {Giles}, \&
  {Bromley}}]{lee2015b}
{Lee}, G., {Helling}, C., {Giles}, H., \& {Bromley}, S.~T. 2015, \aap, 575, A11

\bibitem[{{Lee} {et~al.}(2018){Lee}, {Blecic}, \& {Helling}}]{lee2018}
{Lee}, G.~K.~H., {Blecic}, J., \& {Helling}, C. 2018, Philosophical
  Transactions of the Royal Society of London Series A, arXiv:1801.08482

\bibitem[{{Lee} {et~al.}(2017){Lee}, {Wood}, {Dobbs-Dixon}, {Rice}, \&
  {Helling}}]{lee2017}
{Lee}, G.~K.~H., {Wood}, K., {Dobbs-Dixon}, I., {Rice}, A., \& {Helling}, C.
  2017, \aap, 601, A22

\bibitem[{{Line} \& {Parmentier}(2016)}]{line2016}
{Line}, M.~R., \& {Parmentier}, V. 2016, \apj, 820, 78

\bibitem[{Lodders(1999)}]{lodders1999}
Lodders, K. 1999, The Astrophysical Journal, 519, 793

\bibitem[{{Lodders}(2010)}]{lodders2010}
{Lodders}, K. 2010, Astrophysics and Space Science Proceedings, 16, 379

\bibitem[{{MacDonald} \& {Madhusudhan}(2017)}]{macdonald2017}
{MacDonald}, R.~J., \& {Madhusudhan}, N. 2017, \mnras, 469, 1979

\bibitem[{{Mainzer} {et~al.}(2007){Mainzer}, {Roellig}, {Saumon}, {Marley},
  {Cushing}, {Sloan}, {Kirkpatrick}, {Leggett}, \& {Wilson}}]{mainzer2007}
{Mainzer}, A.~K., {Roellig}, T.~L., {Saumon}, D., {et~al.} 2007, \apj, 662,
  1245

\bibitem[{{Marley} {et~al.}(2010){Marley}, {Saumon}, \&
  {Goldblatt}}]{marley2010}
{Marley}, M.~S., {Saumon}, D., \& {Goldblatt}, C. 2010, \apjl, 723, L117

\bibitem[{McGouldrick \& Toon(2007)}]{mcgouldrick2007}
McGouldrick, K., \& Toon, O.~B. 2007, Icarus, 191, 1

\bibitem[{{Miura} {et~al.}(2010){Miura}, {Tanaka}, {Yamamoto}, {Nakamoto},
  {Yamada}, {Tsukamoto}, \& {Nozawa}}]{miura2010}
{Miura}, H., {Tanaka}, K.~K., {Yamamoto}, T., {et~al.} 2010, \apj, 719, 642

\bibitem[{Morley {et~al.}(2013)Morley, Fortney, Kempton, Marley, Vissher, \&
  Zahnle}]{morley2013}
Morley, C.~V., Fortney, J.~J., Kempton, E. M.-R., {et~al.} 2013, The
  Astrophysical Journal, 775, 33

\bibitem[{Morley {et~al.}(2012)Morley, Fortney, Marley, Visscher, Saumon, \&
  Leggett}]{morley2012}
Morley, C.~V., Fortney, J.~J., Marley, M.~S., {et~al.} 2012, The Astrophysical
  Journal, 756, 172

\bibitem[{Morley {et~al.}(2015)Morley, Fortney, Marley, Zahnle, Line, Kempton,
  Lewis, \& Cahoy}]{morley2015}
---. 2015, The Astrophysical Journal, 815, 110

\bibitem[{Murphy {et~al.}(1993)Murphy, Haberle, Toon, \& Pollack}]{murphy1993}
Murphy, J.~R., Haberle, R.~M., Toon, O.~B., \& Pollack, J.~B. 1993, Journal of
  Geophysical Research, 98, 3197

\bibitem[{{Parmentier} {et~al.}(2016){Parmentier}, {Fortney}, {Showman},
  {Morley}, \& {Marley}}]{parmentier2016}
{Parmentier}, V., {Fortney}, J.~J., {Showman}, A.~P., {Morley}, C., \&
  {Marley}, M.~S. 2016, \apj, 828, 22

\bibitem[{Plane(2012)}]{plane2012}
Plane, J. M.~C. 2012, Chem. Soc. Rev., 41, 6507

\bibitem[{Pruppacher \& Klett(1978)}]{pruppacher1978}
Pruppacher, H.~R., \& Klett, J.~D. 1978, Microphysics of clouds and
  precipitation (D. Reidel Publishing Company, Dordrecht, Holland)

\bibitem[{{Rajan} {et~al.}(2017){Rajan}, {Rameau}, {De Rosa}, {Marley},
  {Graham}, {Macintosh}, {Marois}, {Morley}, {Patience}, {Pueyo}, {Saumon},
  {Ward-Duong}, {Ammons}, {Arriaga}, {Bailey}, {Barman}, {Bulger}, {Burrows},
  {Chilcote}, {Cotten}, {Czekala}, {Doyon}, {Duch{\^e}ne}, {Esposito},
  {Fitzgerald}, {Follette}, {Fortney}, {Goodsell}, {Greenbaum}, {Hibon},
  {Hung}, {Ingraham}, {Johnson-Groh}, {Kalas}, {Konopacky}, {Lafreni{\`e}re},
  {Larkin}, {Maire}, {Marchis}, {Metchev}, {Millar-Blanchaer}, {Morzinski},
  {Nielsen}, {Oppenheimer}, {Palmer}, {Patel}, {Perrin}, {Poyneer},
  {Rantakyr{\"o}}, {Ruffio}, {Savransky}, {Schneider}, {Sivaramakrishnan},
  {Song}, {Soummer}, {Thomas}, {Vasisht}, {Wallace}, {Wang}, {Wiktorowicz}, \&
  {Wolff}}]{rajan2017}
{Rajan}, A., {Rameau}, J., {De Rosa}, R.~J., {et~al.} 2017, \aj, 154, 10

\bibitem[{Saumon \& Marley(2008)}]{saumon2008}
Saumon, D., \& Marley, M.~S. 2008, The Astrophysical Journal, 689, 1327

\bibitem[{Seki \& Hasegawa(1983)}]{seki1983}
Seki, J., \& Hasegawa, H. 1983, Astrophysics and Space Science, 94, 177

\bibitem[{{Sing} {et~al.}(2016){Sing}, {Fortney}, {Nikolov}, {Wakeford},
  {Kataria}, {Evans}, {Aigrain}, {Ballester}, {Burrows}, {Deming},
  {D{\'e}sert}, {Gibson}, {Henry}, {Huitson}, {Knutson}, {Etangs}, {Pont},
  {Showman}, {Vidal-Madjar}, {Williamson}, \& {Wilson}}]{sing2016}
{Sing}, D.~K., {Fortney}, J.~J., {Nikolov}, N., {et~al.} 2016, Nature, 529, 59

\bibitem[{{Skemer} {et~al.}(2016){Skemer}, {Morley}, {Zimmerman}, {Skrutskie},
  {Leisenring}, {Buenzli}, {Bonnefoy}, {Bailey}, {Hinz}, {Defr{\'e}re},
  {Esposito}, {Apai}, {Biller}, {Brandner}, {Close}, {Crepp}, {De Rosa},
  {Desidera}, {Eisner}, {Fortney}, {Freedman}, {Henning}, {Hofmann},
  {Kopytova}, {Lupu}, {Maire}, {Males}, {Marley}, {Morzinski}, {Oza},
  {Patience}, {Rajan}, {Rieke}, {Schertl}, {Schlieder}, {Stone}, {Su}, {Vaz},
  {Visscher}, {Ward-Duong}, {Weigelt}, \& {Woodward}}]{skemer2016}
{Skemer}, A.~J., {Morley}, C.~V., {Zimmerman}, N.~T., {et~al.} 2016, \apj, 817,
  166

\bibitem[{{Stephens} {et~al.}(2009){Stephens}, {Leggett}, {Cushing}, {Marley},
  {Saumon}, {Geballe}, {Golimowski}, {Fan}, \& {Noll}}]{stephens2009}
{Stephens}, D.~C., {Leggett}, S.~K., {Cushing}, M.~C., {et~al.} 2009, \apj,
  702, 154

\bibitem[{{Stevenson} {et~al.}(2017){Stevenson}, {Line}, {Bean}, {D{\'e}sert},
  {Fortney}, {Showman}, {Kataria}, {Kreidberg}, \& {Feng}}]{stevenson2017}
{Stevenson}, K.~B., {Line}, M.~R., {Bean}, J.~L., {et~al.} 2017, \aj, 153, 68

\bibitem[{Suhasaria {et~al.}(2015)Suhasaria, Thrower, \&
  Zacharias}]{suhasaria2015}
Suhasaria, T., Thrower, J.~D., \& Zacharias, H. 2015, Monthly Notices of the
  Royal Astronomical Society, 454, 3317

\bibitem[{Suhasaria {et~al.}(2017)Suhasaria, Thrower, \&
  Zacharias}]{suhasaria2017}
---. 2017, Monthly Notices of the Royal Astronomical Society, stx1965

\bibitem[{Toon {et~al.}(1992)Toon, McKay, Griffith, \& Turco}]{toon1992}
Toon, O.~B., McKay, C.~P., Griffith, C.~A., \& Turco, R.~P. 1992, Icarus, 95,
  24

\bibitem[{Toon {et~al.}(1979)Toon, Turco, Hamill, Kiang, \& Whitten}]{toon1979}
Toon, O.~B., Turco, R.~P., Hamill, P., Kiang, C.~S., \& Whitten, R.~C. 1979,
  Journal of the Atmospheric Sciences, 36, 718

\bibitem[{Toon {et~al.}(1989)Toon, Turco, Jordan, Goodman, \& Ferry}]{toon1989}
Toon, O.~B., Turco, R.~P., Jordan, J., Goodman, J., \& Ferry, G. 1989, Journal
  of Geophysical Research, 94, 11359

\bibitem[{Toon {et~al.}(1988)Toon, Turco, Westphal, Malone, \& Liu}]{toon1988}
Toon, O.~B., Turco, R.~P., Westphal, D., Malone, R., \& Liu, M.~S. 1988,
  Journal of the Atmospheric Sciences, 45, 2123

\bibitem[{{Tsuji}(2002)}]{tsuji2002}
{Tsuji}, T. 2002, \apj, 575, 264

\bibitem[{{Tsuji} {et~al.}(1996){Tsuji}, {Ohnaka}, \& {Aoki}}]{tsuji1996}
{Tsuji}, T., {Ohnaka}, K., \& {Aoki}, W. 1996, \aap, 305, L1

\bibitem[{Turco {et~al.}(1979)Turco, Hamill, Toon, Whitten, \&
  Kiang}]{turco1979}
Turco, R.~P., Hamill, P., Toon, O.~B., Whitten, R.~C., \& Kiang, C.~S. 1979,
  Journal of the Atmospheric Sciences, 36, 699

\bibitem[{{Wakeford} {et~al.}(2017){Wakeford}, {Visscher}, {Lewis}, {Kataria},
  {Marley}, {Fortney}, \& {Mandell}}]{wakeford2017}
{Wakeford}, H.~R., {Visscher}, C., {Lewis}, N.~K., {et~al.} 2017, \mnras, 464,
  4247

\bibitem[{Westwood \& Hitch(1963)}]{westwood1963}
Westwood, A. R.~C., \& Hitch, T.~T. 1963, Journal of Applied Physics, 34, 3085

\bibitem[{{Witte} {et~al.}(2011){Witte}, {Helling}, {Barman}, {Heidrich}, \&
  {Hauschildt}}]{witte2011}
{Witte}, S., {Helling}, C., {Barman}, T., {Heidrich}, N., \& {Hauschildt},
  P.~H. 2011, \aap, 529, A44

\bibitem[{Wolf \& Toon(2010)}]{wolf2010}
Wolf, E.~T., \& Toon, O.~B. 2010, Science, 328, 1266

\bibitem[{Yuan \& Lee(2013)}]{yuan2013}
Yuan, Y., \& Lee, T.~R. 2013, Contact Angle and Wetting Properties, ed.
  G.~Bracco \& B.~Holst (Berlin, Heidelberg: Springer Berlin Heidelberg), 3--34

\bibitem[{Zhao {et~al.}(1995)Zhao, Turco, \& Toon}]{zhao1995}
Zhao, J., Turco, R.~P., \& Toon, O.~B. 1995, Journal of Geophysical Research,
  100, 7315

\end{thebibliography}
\bibliographystyle{aasjournal}

\end{document}